\documentclass[prl,twocolumn,letterpaper, superscriptaddress]{revtex4-1}
\usepackage{graphicx}
\usepackage{dcolumn}
\usepackage{bm}
\usepackage{color}
\usepackage{tabularx}
\usepackage{array}
\usepackage{amsmath}
\usepackage{multirow}
\usepackage{stmaryrd}  

\begin{document}

\title{Simultaneous Optical and Electrical Spin-Torque Magnetometry with Stroboscopic Detection of Spin-Precession Phase}

\author{Yi Li}
\affiliation{Department of Physics, Oakland University, Rochester, MI 48309, USA}
\affiliation{Materials Science Division, Argonne National Laboratory, Argonne, IL 60439, USA}

\author{Hilal Saglam}
\affiliation{Materials Science Division, Argonne National Laboratory, Argonne, IL 60439, USA}
\affiliation{Department of Physics, Illinois Institute of Technology, Chicago IL 60616, USA}

\author{Zhizhi Zhang}
\affiliation{Materials Science Division, Argonne National Laboratory, Argonne, IL 60439, USA}
\affiliation{School of Optical and Electronic Information, Huazhong University of Science and Technology, Wuhan 430074, China}

\author{Rao Bidthanapally}
\affiliation{Department of Physics, Oakland University, Rochester, MI 48309, USA}

\author{Yuzan Xiong}
\affiliation{Department of Electrical and Computer Engineering, Oakland University, Rochester, MI 48309, USA}

\author{John E. Pearson}
\affiliation{Materials Science Division, Argonne National Laboratory, Argonne, IL 60439, USA}

\author{Valentine Novosad}
\affiliation{Materials Science Division, Argonne National Laboratory, Argonne, IL 60439, USA}

\author{Hongwei Qu}
\affiliation{Department of Electrical and Computer Engineering, Oakland University, Rochester, MI 48309, USA}

\author{Gopalan Srinivasan}
\affiliation{Department of Physics, Oakland University, Rochester, MI 48309, USA}

\author{Axel Hoffmann}
\email{hoffmann@anl.gov}
\affiliation{Materials Science Division, Argonne National Laboratory, Argonne, IL 60439, USA}

\author{Wei Zhang}
\email{weizhang@oakland.edu}
\affiliation{Department of Physics, Oakland University, Rochester, MI 48309, USA}
\affiliation{Materials Science Division, Argonne National Laboratory, Argonne, IL 60439, USA}

\date{\today}

\begin{abstract}

Spin-based coherent information processing and encoding utilize the precession phase of spins in magnetic materials. However, the detection and manipulation of spin precession phases remain a major challenge for advanced spintronic functionalities. By using simultaneous electrical and optical detection, we demonstrate the direct measurement of the precession phase of Permalloy ferromagnetic resonance driven by the spin-orbit torques from adjacent heavy metals. The spin Hall angle of the heavy metals can be independently determined from concurrent electrical and optical signals. The stroboscopic optical detection also allows spatially measuring local spin-torque parameters and the induced ferromagnetic resonance with comprehensive amplitude and phase information. Our study offers a route towards future advanced characterizations of spin-torque oscillators, magnonic circuits, and tunnelling junctions, where measuring the current-induced spin dynamics of individual nanomagnets are required.

\end{abstract}

\maketitle

\section{Introduction}

Recent breakthroughs in spin information transport bring about new paradigms for spintronic information processing \cite{HoffmannPRApplied2015,CornelissenNPhys2015,WesenbergNPhys2017,LebrunNature2018}, where information is carried by magnetic excitations, or magnons. While most research focuses on how to maintain high signal amplitude, its phase manipulation has also received increased attentions in the area of magnonics \cite{KruglyakJPD2010,PerzlmaierPRB2008}, spin wave logics \cite{SchneiderAPL2008} and synchronized dynamics in spin-torque oscillators \cite{KakaNature2005,MancoffNature2005,LiPRL2017}. Furthermore, in analogy to quantum information, magnons have been demonstrated to coherently couple with magnetization dynamics from other magnon sources \cite{KlinglerPRL2018,ChenPRL2018}, electromagnetic waves (photons) \cite{HueblPRL2013,TabuchiPRL2014,ZhangPRL2014,BaiPRL2015}, and lattice vibrations (phonons) \cite{ZhangScienceAdv2016,KikkawaPRL2016}, for achieving hybridized dynamics. For such applications, the capability to accurately measure and tune the phase of local magnetic excitations is paramount.

Spin-orbit torques (SOTs) provide a unique pathway for manipulating both the amplitude and the phase of magnetization precession \cite{SklenarSpintronicsXI2018,SklenarPRB2017Unidirectional}. The SOTs can be generated via the spin Hall effect (SHE), by applying a charge current through an adjacent heavy metal layer, which injects due to spin-orbit coupling a pure spin current to the ferromagnet (FM) \cite{SinovaRMP2015}. The longitudinal SOT, also known as the anti-damping torque, has a symmetry of $ \textbf{m} \times \sigma \times \textbf{m} $ (where $\textbf{m}$ is the magnetization vector and $\sigma$ is the spin polarization of the injected spin current), and thus a 90-degree phase difference compared with torques from the Oersted field and the field-like torque ($ \sigma \times \textbf{m} $) generated from a charge current. The phase difference between the magnetization precession and the driving current have been measured via electrical rectification signals in order to quantify the spin Hall angle, $\theta_{SH}$, the parameter dictating the strength of the spin Hall effect of the materials \cite{SankeyPRL2006,MosendzPRL2010,LiuPRL2011,WeilerPRL2014}. However, such an electrical means only access to the spin dynamics "indirectly" via the rectification mechanism, and parasitic electric effects often emerge such as the inverse spin Hall signal \cite{BaiLHPRL2013,HarderPRB2011} and the propagation delay of electromagnetic wave \cite{BaileyNcomm2013}, which can complicate the lineshape analysis. Besides, electrical measurements also require an additional readout circuit and cannot easily provide spatial resolution of the device components in a complicated microwave circuit \cite{VlaminckAPL2012}.

Magneto-optical Kerr effect (MOKE) provides an alternative approach to ``directly" access to the magnetization states. Both in-plane and out-of-plane magnetic moments can be detected \cite{BaderRSI2000} by the choice of different Kerr configurations. Recently, MOKE-based detection of electrically-induced SOTs have been reported \cite{FanAPL2016,HayashiAPEX2018,TsaiSREP2018, MontazeriNcomm2015}, but only with quasi-static magnetization configurations, where the SOT is treated as an "effective field" that tilts the static magnetization of the FM. On the other hand, stroboscopic techniques \cite{nembachPRL2013,MoriyamaJAP2015,GuoPRAppl2015,FuchsNcomm2015,YoonPRB2016}, in which both the pump and probe are modulated at the dynamic excitation frequency, offer unique advantages in tracking both the amplitude and the phase information of the magnetization precession, and thus are more suitable in studying SOT-driven spin dynamics.

\begin{figure*}[htb]
 \centering
 \includegraphics[width=5.6 in]{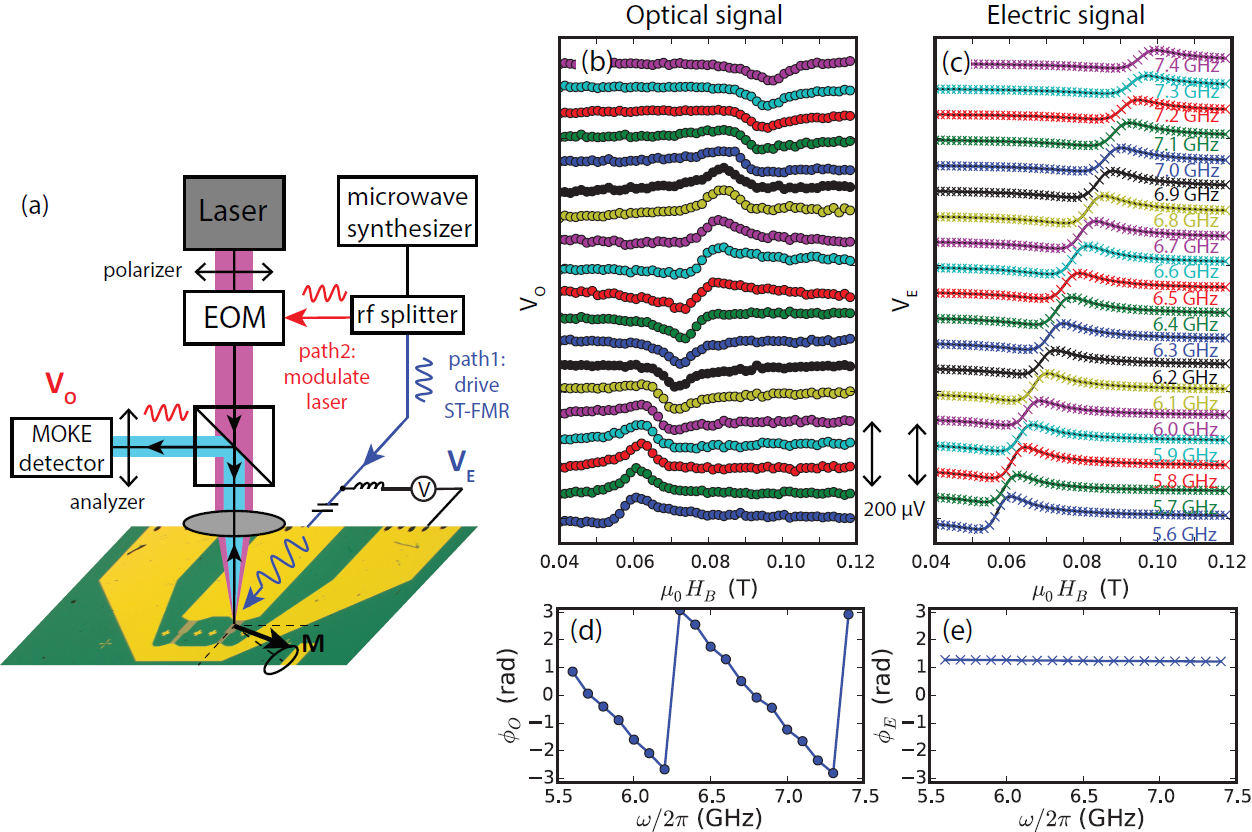}
 \caption{(a) Illustration of the microwave excitations and optical measurements of spin-torque FMR using our combined electrical and optical technique. (b-c) Optical and electrical signals of spin-torque FMR spectra for Pt(6 nm)/Py(6 nm) bilayer devices from 5.6 GHz to 7.4 GHz. (d-e) Extracted optical and electrical phases, $\phi_O$ and $\phi_E$, from Eq. (\ref{eq01}).}
 \label{fig1}
\end{figure*}

Here, we report a simultaneous electrical and optical measurement of spin-torque ferromagnetic resonance (ST-FMR), with the capability to extract the spin precession phase driven by the SOTs. {\color{black}We show that the spin Hall angle of heavy metals can be directly extracted from the measured optical phase of spin dynamics, independently from the electrically-detected ST-FMR. Using this method and as a demonstration, we measure the spin Hall angle for Pt, $\sim$ 0.052$\pm0.009$, and Ta, $\sim$ -0.034$\pm0.021$. These values are also in agreement with the concurrent and independent electrical measurements. Furthermore, the optically measured phase is insensitive to the ST-FMR device configuration, and thus provides a robust characterization of the heavy metal generating the SOT. Moreover, we show that such optical technique allows us to gain spatial resolution of the amplitude and phase of the spin dynamics in microstructured devices. }

\section{Results and discussions}

\textbf{Phase evolution of magnetization dynamics.} We fabricate ST-FMR devices on Si/SiO$_2$ substrates by magnetron sputtering and optical lithography. The  devices consist of nonmagnetic metal (NM = Pt, Ta, and Cu) / Ni$_{80}$Fe$_{20}$ (Permalloy, Py) bilayers, with the Py layer (6 nm) on top for optical access. The dimension of the bilayer is 100 $\mu m$ $\times$ 400 $\mu m$, and the thicknesses for the NM layers are Cu (10 nm), Pt (6 nm), and Ta (6 nm). As references, we also prepared devices of Pt/SiO$_2$/Py and Ta/SiO$_2$/Py trilayers, with a thin SiO$_2$ (1.5 nm) spacer in order to suppress the interfacial transmission of the spin current \cite{MosendzAPL2010}. The thin SiO$_2$ layer also does not alter the current flow pattern significantly. The simultaneous electrical and optical detection of the ST-FMR is schematically shown in Fig. \ref{fig1}(a). From a microwave splitter, path 1 is used to continuously drive the magnetization dynamics; path 2 is used to modulate the laser for stroboscopic measurements. To reduce the white noise and also enable heterodyne detection, a low-frequency signal of $f_\text{mix}=100$ kHz is mixed to path 1 and a Kerr signal $V_{O}$ at the same frequency is detected in path 2 by a lock-in amplifier \cite{YoonPRB2016}. At the same time, the electrical rectification signal $V_{E}$ is also recorded in path 1 by a dc nanovoltmeter. The detailed measurement setup and components are summarized in the Supplemental Material.

Figs. \ref{fig1}(b-c) compare the simultaneously measured $V_{O}$ and $V_{E}$ at different microwave frequencies $\omega$, for a Pt/Py device. For both cases, the lineshapes can be fitted to a complex Lorentzian function:
\begin{equation}
V_{E,O} = \operatorname{Re}\left[{A_{E,O}  e^{i\phi_{E,O}} \over (H_{B}-H_{res})+i\Delta H_{1/2}/2}\right],
\label{eq01}
\end{equation}
where $H_{B}$ is the DC biasing field, $H_{res}$ is the resonance field, $\Delta H_{1/2}$ is the full-width-half-maximum linewidth, and $A$ is the amplitude. The extracted phase $\phi$, which contains the information of the magnetization precession phase, mixes the real and imaginary parts of the denominator, which corresponds to asymmetric and symmetric Lorentzian lineshapes, respectively \cite{MosendzPRL2010,LiuPRL2011}.

The extracted phase from optical ($\phi_{O}$) and electrical ($\phi_{E}$) signals differ significantly, which are shown in Figs. \ref{fig1}(d-e). Because $V_{O}$ is rectified from an independent laser path, $\phi_{O}$ is a direct reflection of the precession phase of Py, whereas $\phi_{E}$ represents the phase difference between the local microwave current and magnetization motion with the phase of the current dependent on the electrical path. For $\phi_{O}$, its frequency dependence is caused by the light path difference $\Delta L$ between path 1 \& 2, following $\phi_{O} = \phi_{O}(0)+\omega \Delta L/c$. From the slope in Fig. \ref{fig1}(d) we calculated $\Delta L \sim 30$ cm. For $\phi_{E}$ in a conventional ST-FMR analysis \cite{LiuPRL2011}, it represents the spin Hall angle of Pt.

\begin{figure}[b]
 \centering
 \includegraphics[width=3.4 in]{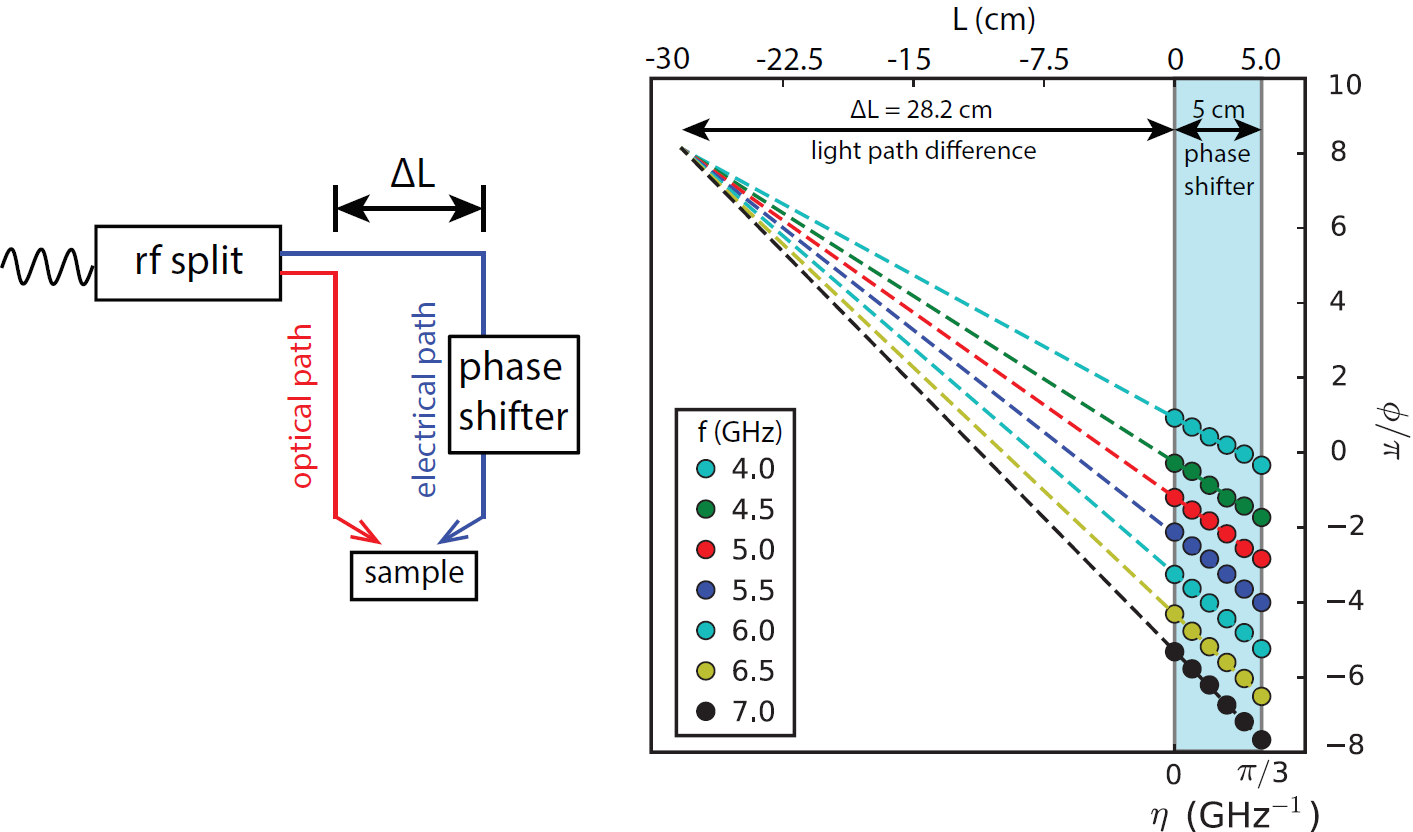}
 \caption{Left: introduction of the phase shifter to the electrical path. Right: evolution of $\phi_O$ measured on Pt(6 nm)/Py(6 nm) device at different phase tuning ranges $\eta$ and frequencies $\omega$. Dashed lines are linear fits to the data. }
 \label{fig2}
\end{figure}

\textbf{Additional phase control with a delay line.} The precession phase of Py can be tuned not only by changing the driving frequency, but also by changing the length of the delay line. In Fig. \ref{fig2}, we introduce a phase shifter to the electrical path, which allows very fine tuning of the delay path length. The fine tuning range $\eta$ is from zero to $\pi/3$ GHz$^{-1}$ which is equivalent to 5 centimeters of delay line. The extracted $\phi_O$ are plotted in Fig. \ref{fig2} for different $\eta$ and $\omega$. Here, we restrict $\phi_O$ of $\omega/2\pi=4$ GHz between $-\pi$ and $\pi$ and add offsets to all measured $\phi_O$ so that it is continuous, allowing a grand linear fit. By linearly extrapolating $\phi_O$-$\eta$ for different $\omega$, all frequencies cross each other at the same point, which corresponds to a zero length difference between the optical and electrical paths. From this set of data we also confirm $\Delta L=28.2$ cm, which agrees with the value from the frequency-dependent phase variation.

\begin{figure}[htb]
 \centering
 \includegraphics[width=3.0 in]{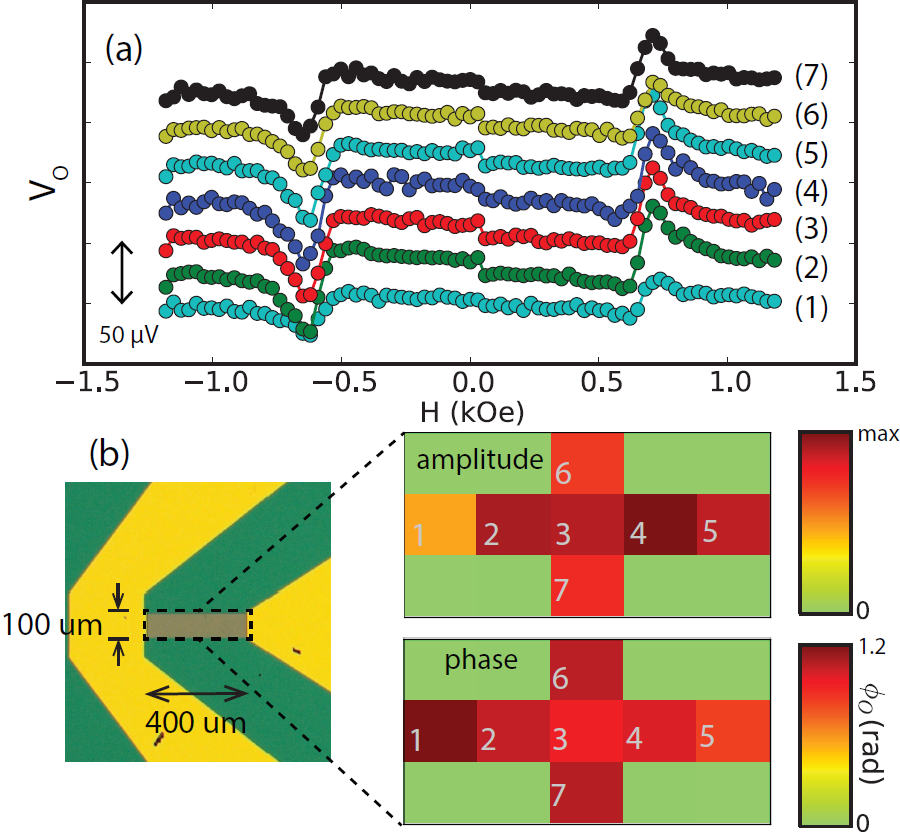}
 \caption{(a) Lineshapes of optical signals of Pt(6 nm)/Py(6 nm) device at difference locations. (b) Illustration of the spatial MOKE detection, with the extracted amplitudes and phases shown in color diagrams. }
 \label{fig3}
\end{figure}

\textbf{Spatial resolution of the precession amplitude and phase.} In microstructured spin-torque devices, the uniformity and coherence of spin dynamics has been a long-standing issue, where the central, uniform mode are often undermined by the localized solitons and standing spin wave edge modes \cite{JungfleischPRL2016,LiNanoscale2016}. To verify the uniformity of spin precession in our ST-FMR devices, we measure the optical signals at selective, representative positions of the device for $\omega/2\pi=4$ GHz. Their lineshapes, extracted amplitudes $A$ and phases $\phi_{O}$ are compared in Fig. \ref{fig3}. Across the 400 $\mu$m device bar the value of $\phi_{O}$ stays almost constant within a maximal variation of 0.2 rad. This is in consistent with the negligible phase delay of the microwave along the device, calculated to be $\sim$ 0.04 rad, so that the whole device is considered driven uniformly at the same microwave phase. We also note that the present spatial resolution is limited only by the spherical lens used in this experiment, and it should not reflect the general technical limit of this method. Higher spatial resolution and raster scanning capabilities should be generally possible if one uses a microscopy objective lens and a scanning lens.

\begin{figure*}[htb]
 \centering
 \includegraphics[width=4.0 in]{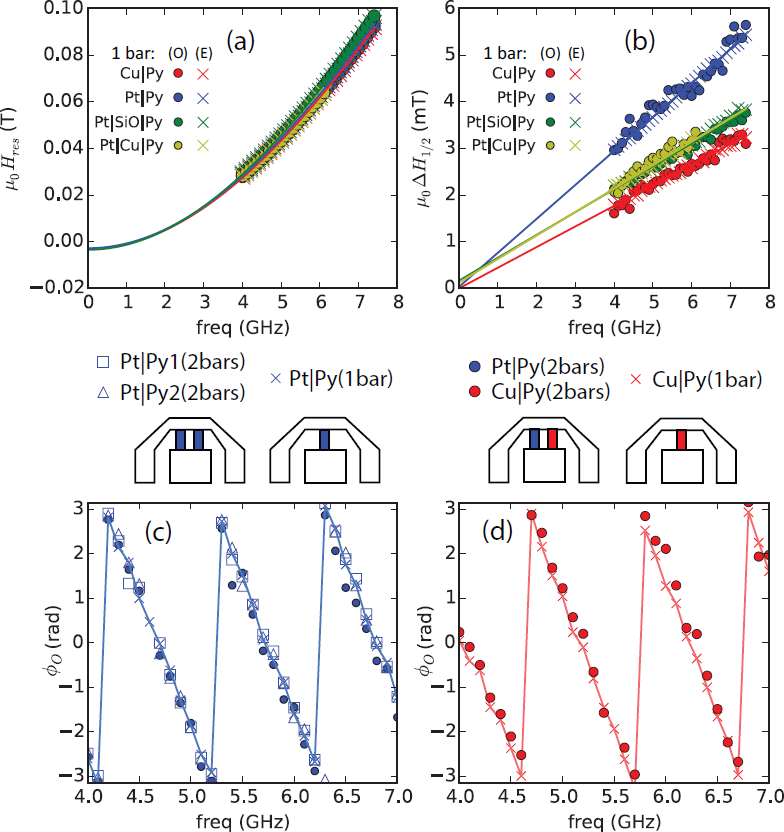}
 \caption{(a-b) Extracted $H_{res}$ and $\Delta H_{1/2}$ values and the corresponding curve fittings of both electrical (crosses) and optical (dots) signals for the single- and dual-bar devices. (c-d) Extracted $\phi_O$ for (c) Pt(6 nm)/Py(6 nm) and (d) Cu(10 nm)/Py(6 nm) devices from the two single-bar samples and the two dual-bar samples. Solid curves are guides to the eye.}
 \label{fig4}
\end{figure*}

\textbf{Simultaneous electrical and optical detection.} Next, we examine and analyze the optical signals in parallel with the electrical counterparts. Figs. \ref{fig4}(a-b) show the extracted $H_{res}$ and $\Delta H_{1/2}$ for Cu/Py, Pt/Py, Pt/SiO$_2$/Py and Pt/Cu/Py devices. The good agreement between electrical (crosses) and optical (circles) measurements indicates high reliability and reproducibility of the optical measurements for probing local magnetization dynamics, Fig. \ref{fig4}(a). Only slightly added noise level is observed in the optical data for $\Delta H_{1/2}$, which can be likely improved further by using a high-end photodetector. In such a series of samples, the linewidth differences reflect the different spin pumping with Pt and the shunting effect with Cu interlayer, Fig. \ref{fig4}(b). The linewidth evolution (from large to small) follows: Pt/Py $>$ Pt/Cu/Py $>$ Pt/SiO$_2$/Py $>$ Cu/Py. In Fig. \ref{fig4}(a), the solid curves are the fits to the Kittel equation $\omega^2/\gamma^2=\mu_0^2(H_{res}+H_k)(H_{res}+H_k+M_{eff})$, and in Fig. \ref{fig4}(b), the Gilbert-type linewidth is fitted with $\mu_0\Delta H_{1/2} = \mu_0\Delta H_0 + 2\alpha\omega/\gamma$, where $\gamma = 2\pi(g_{eff}/2)\cdot 28$ GHz/T and $g_{eff}$ is taken as 2.06, $M_{eff}$ is the effective magnetization, $H_k$ is the anisotropy field, $\alpha$ is the Gilbert damping and $\Delta H_0$ is the inhomogeneous linewidth. From the fitting, we find the effective magnetization, $\mu_0M_{eff}=0.60, 0.59$ and $0.56$ T for Cu/Py, Pt/Py and Ta/Py, respectively, and the fitting results of $\alpha$ and spin mixing conductance  $g^{\uparrow\downarrow}$ \cite{DuCHPRApplied2014} are shown in Table I.

We highlight the advantage of the spatial selectivity in the optical measurements by fabricating and measuring additional, dual-bar devices as shown in Fig. \ref{fig4}. In such dual-bar structures, one device consists of two identical Pt/Py stripes and the other consists of one Pt/Py and one Cu/Py stripe. Their phases are compared with the two individual Pt/Py and Cu/Py devices. In Fig. \ref{fig4}(c) and (d), the extracted $\phi_{O}$ for the total six stripes are plotted. We found that the phases are in good agreement for all four Pt/Py stripes, Fig. \ref{fig4}(c), as well as for both two Cu/Py stripes, Fig. \ref{fig4}(d). Such information cannot be achieved by electrical detections, where the signal would only come from the averaged rectification voltage from the two tripes. Our results also show the repeatability and low systematic errors for determining the precession phase for difference devices, which is crucial for elucidating the spin-torque induced precession phase variation in the following section. We also find a large phase offset between Pt/Py and Cu/Py. This is mainly due to the surface/interface reflectivity, which will be discussed later.

\begin{figure*}[htb]
 \centering
 \includegraphics[width=6.7 in]{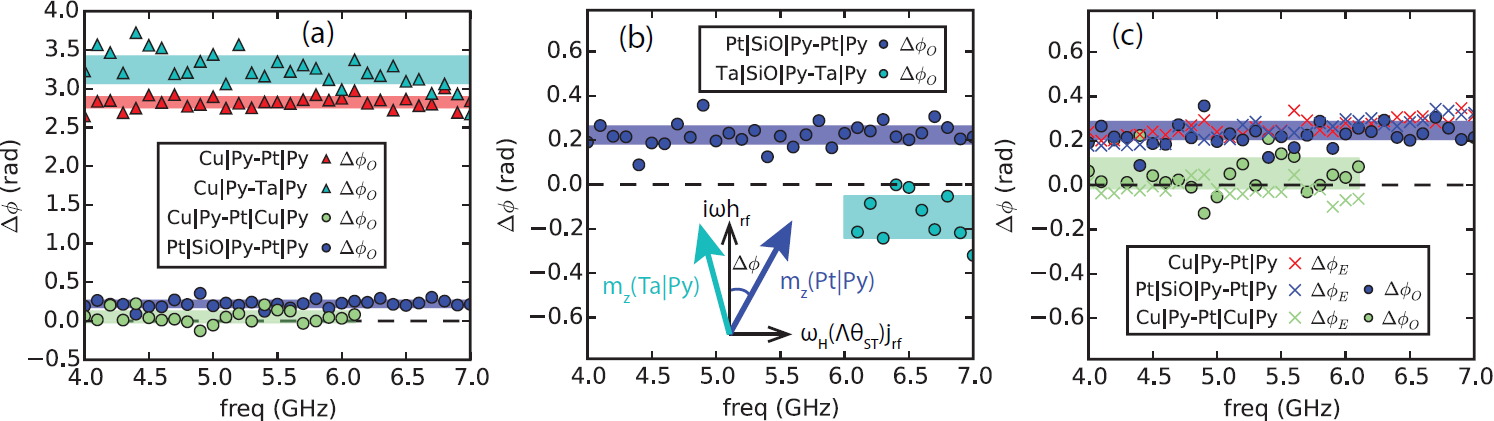}
 \caption{(a) Comparison of the optical phase offset for different reflection interfaces (triangles). The shaded areas represent the mean values and errorbars of the measurement data. (b) $\Delta \phi_{O}$ induced by Pt (blue) and Ta (cyan). The inset shows the relation of the additional, spin-torque induced precession phases with respect to the $h_{rf}$. (c) Comparison of the $\Delta \phi_{O}$ and $\Delta \phi_{E}$ in the Pt series of samples.}
 \label{fig5}
\end{figure*}

\begin{table*}[htb]
\centering
\setlength{\extrarowheight}{1.5pt}
\begin{tabular}{>{\centering\arraybackslash}m{0.4in} | >{\centering\arraybackslash}m{0.4in} >{\centering\arraybackslash}m{0.5in} >{\centering\arraybackslash}m{0.5in} >{\centering\arraybackslash}m{0.5in} >{\centering\arraybackslash}m{0.5in} >{\centering\arraybackslash}m{0.8in} >{\centering\arraybackslash}m{0.8in} }
\hline
\hline
 &R &R(+SiO$_2$)&$\alpha$ &$\alpha$(+SiO$_2$) &$g^{\uparrow\downarrow}$ & $\Delta\phi_O$ & $\theta_{SH}$ \\
 &($\Omega$) &($\Omega$)& &  &(nm$^{-2}$) & (rad)   &  \\
\hline
Cu/Py & 26.9 & -  & 0.0064 & -      &- & - & - \\
Pt/Py & 102.6 & 174.8 & 0.0105 & 0.0071 &10.7 & $0.22\pm0.04$  & $0.052\pm0.009$ \\
Ta/Py & 316.9 & 177.0 & 0.0082 & 0.0068 &4.4 & $-0.15\pm0.09$  & $-0.034\pm0.021$ \\
\hline
\end{tabular}
\caption{Summary of the device resistance, Gilbert damping, spin-mixing conductance, electrical and optical phases, and the spin Hall angles.}
\label{table1}
\end{table*}

\textbf{Spin-torque effects on the precession phase and determination of the spin Hall angle.} We now shift our focus to the key results of this work, which is the demonstration of the spin-torque manipulation on the spin precession phase. We first note our observations of a finite, offset phase caused by the different NM underlayers for Py. In the most ideal scenario, the probing laser beam should reflect just from the Py surface, and should not penetrate across the NM / Py interface. However, in reality, the laser penetration depth is estimated $\sim$ 10 nm, exceeding the top Py thickness (6 nm) for all samples. Therefore, the different materials under the Py layer introduce a nontrivial, relative phase offset, which is solely due to the magneto-optics, and is irrelevant to the spin-torque effects. Such an offset phase is robust for the different NM underlayer materials and is negligible for SiO$_2$ interlayers (1.5 nm). We compare the magneto-optics induced phase shifts between different samples in Fig. \ref{fig5}(a). For example, large $\Delta\phi_{O}$ close to $\pi$ is observed when the Cu layer (in Cu/Py) is replaced by Pt (in Pt/Py) or Ta (in Ta/Pt). However, since the laser penetration is $\sim$ 10 nm, having Pt underneath the 10-nm Cu layer in the trilayer structures (substrate/Pt/Cu/Py) cause negligible phase offset. Physically, such phase offset due to magneto-optics can be attributed to the variation of refraction index in Cu, Pt and Ta \cite{BaderRSI2000,ChoiPRB2015,ChoiNaturePhys2015}.

The extraction of spin-torque phases are achieved by comparing the Pt(Ta)/Py with Pt(Ta)/SiO$_2$/Py, as shown in Fig. \ref{fig5}(b), where the differences in the optical phase, $\Delta\phi_{O}$, should be caused solely by spin-orbit torques from Pt(Ta). We found a "+" optical phase,  $\Delta\phi_O = 0.22 \pm 0.04 $ rad, induced by Pt, and a "-" optical phase, $\Delta\phi_O = -0.15 \pm 0.09 $ rad, induced by Ta, respectively. In particular, the +(-) phase induced by Pt(Ta) indicates a phase-lag(advance) of the magnetization vector caused by the corresponding anti-damping torques, shown by the inset of Fig. \ref{fig5}(b). Therefore, our result provides a direct and sensitive access to the spin-torque effects on the spin precession motion, free from any electrical parasitic effects. Further, such phase values are in good agreement with the reversed orientation of the anti-damping torques and thus the opposite spin Hall angles of Pt and Ta, obtained previously using electrical means \cite{MosendzPRL2010,LiuScience2012,HahnPRB2013,WangPRL2014}.

The obtained spin-torque phases also allow a direct extraction of the spin Hall angle of the NMs. To gain a more quantitative interpretation, we show that the perpendicular magnetization, $m_y$, can be derived from the Landau-Lifshitz-Gilbert equation, as:
\begin{equation}
\tilde{m_y}  \sim  {i\omega\tilde{h}_{rf}+\omega_H (\Lambda\theta_{SH})\tilde{j}_{rf} \over (H_B - H_{res})+i\alpha\omega/\gamma}
\label{eq02}
\end{equation}
where $\omega_H=\gamma\mu_0H_B$, $\tilde{h}_{rf}=\tilde{j}_{rf}d_{NM}/2$ is the microwave field acting on the magnetization, $\tilde{j}_{rf}$ and $d_{NM}$ are the current density and thickness of the NM layer, respectively. For the spin-torque-related components, $\theta_{SH}$ is the spin Hall angle, and the spin-charge conversion coefficient, $\Lambda=\hbar/2e\mu_0M_st_{Py}$, shows the effective field for a given spin current \cite{slonczewskiJMMM1996}, with $t_{Py}$ denoting the thickness of Py. Here, we note that the Eq. (\ref{eq02}) is also the physical representation of Eq. (\ref{eq01}).

As shown by the inset of Fig. \ref{fig5}(b), the $i\omega\tilde{h}_{rf}$ and $\omega_H (\Lambda\theta_{SH})\tilde{j}_{rf}$ represent the Oersted field and the anti-damping torque induced by the spin Hall effect, respectively. The phase shift can be expressed as \cite{LiuPRL2011,WeiZhangPRB2015}:
\begin{equation}
\tan\Delta\phi = {\omega_H(\Lambda\theta_{SH})\tilde{j}_{rf} \over \omega \tilde{h}_{rf}} = {\omega_H\over \omega} {2\Lambda\theta_{SH} \over d_{NM}}
\label{eq03}
\end{equation}
where at the resonance condition, $\omega_H/\omega=1/\sqrt{1+M_{eff}/H_{res}}$. The calculated values of $\theta_{SH}$ for Pt and Ta are listed in Table I.

Finally, we also compare the optical data with the simultaneously measured electrical ST-FMR data. As shown in Fig. \ref{fig5}(c), $\Delta\phi_E$ taken by subtracting the phases from Pt/Py and Pt/SiO$_2$/Py (blue crosses) closely match the previously discussed $\Delta\phi_O$ (blue dots). Their coincidence shows the robustness and accuracy of our optical and electrical phase detection, and also confirms that the previously discussed optical phase shift, $\Delta\phi_O$, is purely spin-torque driven. In addition, the $\Delta\phi_E$ taken from subtracting the phases from Cu/Py and Pt/Py (red crosses) confirms that both Cu/Py and Pt/SiO$_2$/Py have zero spin-torque effect on Py. Lastly, when the Pt is separated from Py by a relatively thick Cu layer (10 nm), both the electrical and optical phase shifts are quite small, yielding $\Delta\phi_E=-0.02\pm0.03$ rad (green crosses) and $\Delta\phi_O=0.05\pm0.07$ rad (green circles), due to primarily the electrical shunting effects, which only contributes to a large Oersted field to the Py spin dynamics, and reduces the current through Pt and therefore the concomitant spin-orbit torques.

\section{Conclusion}

The unique advantages of the stroboscopic MOKE detection allow for a direct, sensitive measure of the spin precession phase and other spintronic parameters induced by local spin-orbit torques in ST-FMR devices. Our combinatorial technique will be useful for future characterizations of spin-torque oscillators, magnonic circuits, and tunnelling junctions, where measuring the dynamics of individual nanomagnets are needed \cite{LiPRL2017,nembachPRL2013,LiNanoscale2016,JungfleischPRL2016,GuoPRAppl2015,GuoPRL2013,JiangPRApplied2018,ZhouZY_AM_2018,ZhouZY_NCOMM_2018}.

In addition, it is useful in the scenarios when electrical detection becomes technically challenging, such as for magnetic insulators where direct electrical modulation is unavailable \cite{JungfleischNanoLett2017,SklenarPRB2015,SchreierPRB2015}, and in particular, when studying the spin-swapping, anomalous-Hall, and planar-Hall torques discovered recently in pure ferromagnets \cite{MohamedPRL2016, GibbonsPRApplied2018, AminPRL2018,BaekNatureMater2018}. The stroboscopic method will also be in complementary to Brillouin light scattering \cite{SergaAPL2006} which measures inelastic light scattering by magnon-phonon, and X-ray magnetic circular dichroism \cite{ArenaPRB2006,BaileyNcomm2013,ZQQiuPRL2016}.

For quantifying the spin Hall angles, the stroboscopic optical detection has an advantage over the DC magneto-optical measurements \cite{FanNcomm2014,MontazeriNcomm2015,FanAPL2016,TsaiSREP2018}. In the static Kerr measurements, a large DC charge current needs to be applied in order to generate pronounced DC spin torques for tilting the magnetization, which can lead to heating-induced nonlinear signals \cite{MontazeriNcomm2015,HayashiAPEX2018}. For the stroboscopic detection, a much smaller microwave current will suffice with negligible heating effect. This is because the resonant excitation of magnetization motion is much more efficient than in DC condition by a factor of 1/$\alpha$, which is more than 100 in the case of Py.

We also comment that the spin Hall angle $\theta_{SH}$ obtained in Table 1 are the damping-like contribution from the spin Hall effects. For Pt/Py interface, field-like torques have also been reported due to the Rashba effect \cite{NanPRB2015,PaiPRB2015,BergerPRB2018}, which will be reflected as an imaginary term of $\theta_{SH}$ in the stroboscopic optical detection. However, because the field-like torque is usually much smaller than the Oersted field, its influence on $\Delta\phi$ is negligible, and such a topic is out of the scope of the current work, and worth another series of future investigations.

\section{Acknowledgements}

W.Z. gratefully acknowledges Oakland University Startup Funds for equipment acquisition and measurement setup construction. This Work was supported by the U.S. National Science Foundation under Grants No. DMR-1808892 and the Michigan Space Grant Consortium. Work at Argonne, including thin films synthesis and device fabrications were supported by the U.S. Department of Energy (DOE), Office of Science, Materials Science and Engineering Division. The use of Center for Nanoscale Materials is supported by DOE-BES, under Contract No. DE-AC02-06CH11357. We also acknowledge fruitful discussions with Dr. Xinlin Song and Prof. Vanessa Sih at the University of Michigan.

\newpage

\onecolumngrid

\newpage

\LARGE{\textbf{Supplemental Materials:}}
\section{Simultaneous Optical and Electrical Spin-Torque Magnetometry with Stroboscopic Detection of Spin-Precession Phase}
\normalsize
\textit{by} Yi Li, Hilal Saglam, Zhizhi Zhang, Rao Bidthanapally, Yuzan Xiong, John E. Pearson, Valentine Novosad, Hongwei Qu, Gopalan Srinivasan, Axel Hoffmann, and Wei Zhang
\newline

\renewcommand{\theequation}{S-\arabic{equation}}
\setcounter{equation}{0}  
\renewcommand{\thefigure}{S-\arabic{figure}}
\setcounter{figure}{0}  

\textbf{1. Measurement setup:}

The spin-torque ferromagnetic resonance (ST-FMR) signals are simultaneously detected electrically by spin rectification and optically by magneto-optic Kerr effect (MOKE) which stroboscopically measures the out-of-plane component of precessing magnetization. A heterodyne method is adopted to enable precessional phase extraction using a setup illustrated in Fig. S1. A single microwave source (BNC-845, Berkeley-Nucleonics) was used to simultaneously modulate the detecting laser light (optical path), and drive the FMR of the sample (electrical path). On the optical path, we used a 1550-nm pigtail-fiber laser (Thorlabs LPSC-1550-FC) for generating continuous-wave light with adjustable power up to 5 mW. The laser light was modulated at the microwave source frequency using an electro-optic intensity modulator (EOM, Optilab IM-1550-12-PM). For optimal laser power, we also used a laser amplifier (Thorlabs EDFA100S) and a fiber-based polarization controller (Thorlabs FPC032) in front of the EOM. The modulated laser light was then converted to free space and polarized before being focused onto the sample surface. The focused light spot is set to $\sim$ 40 $\mu$m in this work. The electric path of the measurement is similar to the conventional ST-FMR measurement using a bias-Tee (Mini-Circuits, ZX85-12G-S+) and a nanovoltmeter (Keithley 2182a).

\begin{figure*}[htb]
 \centering
 \includegraphics[width=4.9 in]{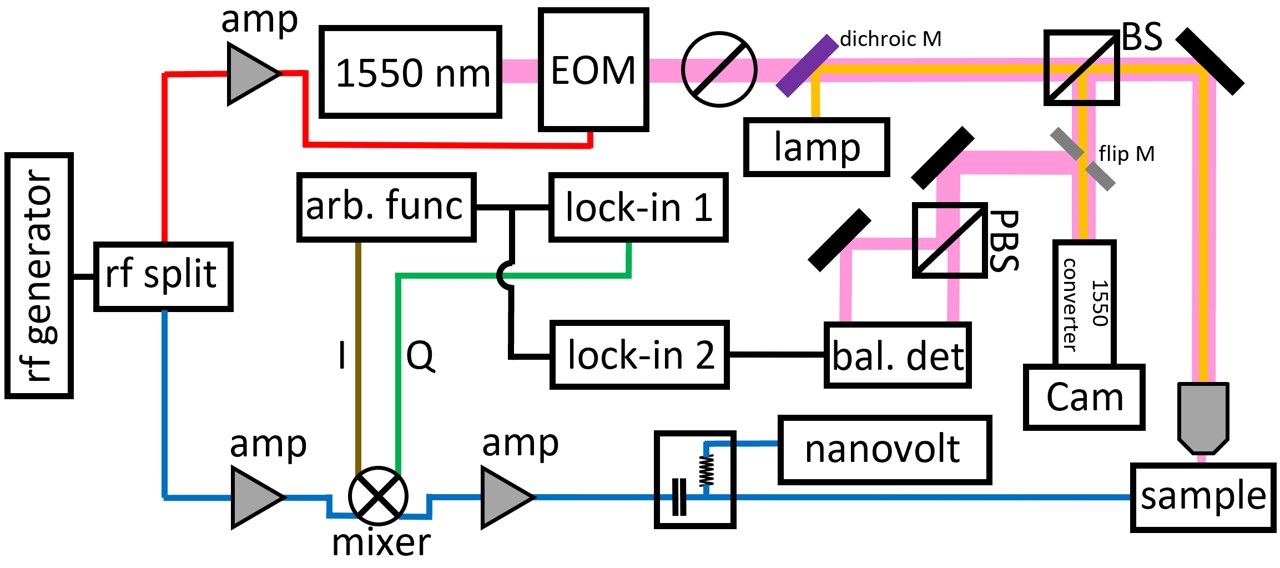}
 \caption{Schematics of the simultaneous electrical and optical-MOKE measurements of ST-FMR. After the rf splitter, the optical path (upper part) contains amplifier, 1550 nm infrared laser module, electro-optic modulator (EOM), polarizer, beam splitter (BS) and focusing lens; the electrical path (lower part) contains amplifier, mixer, bias-Tee, and nanovoltmeter. (PBS = polarizing beam splitter, Cam = camera, bal.det = balancing detector, arb. func = arbitrary waveform generator.)}
 \label{figS1}
\end{figure*}

For a heterodyne detection, the microwave signal along the electrical path was IQ-mixed (Pasternack PE86X9000) with a low-frequency (100 kHz) signal provided by a waveform generator (Keysight 33621A) and a TTL-synchronized, lock-in amplifier (Stanford Research SR830). The voltage amplitude, offset, and phase for the respective "I" and "Q" channels were optimized to ensure the power of the upper side-band of the microwave signal (which was subsequently used for FMR excitation) far exceeds those of the central and lower side-band ($>$20 dB). In this experiment, we used 430 mVpp outputs from the waveform generator and the lock-in, and a 98$^{\circ}$ phase difference between the "I" and "Q" for optimal side-band performance, monitored simultaneously by a real-time spectrum analyzer (RTSA-7550, Berkeley-Nucleonics) and a 6-GHz digital oscilloscope (Keysight DSOX6002A).

The resultant, out-of-plane, dynamical Kerr response of the sample was then probed by the modulated light, sent into a balancing detector (Thorlabs PDB210C) after polarization splitting (Thorlabs PBS254), and analyzed by another lock-in amplifer (Stanford Research SR830). A series of rf amplifiers (Mini-circuits,  ZX60-8008E-S+, ZX60-14012L-S+) and programmable attenuators (RUDAT-13G-60) were also used for signal strength adjustments as needed. The device synchronization and software control were conveniently achieved by customized modules and central programming interfaces developed by THATec Innovation GmbH. \\

\textbf{2. Spin-torque driven ferromagnetic resonance formalism:}

\begin{figure}[htb]
 \centering
 \includegraphics[width=2.5 in]{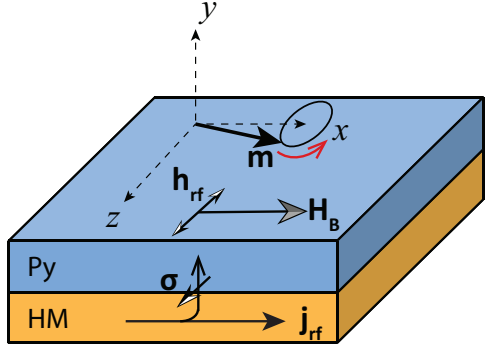}
 \caption{Vector diagram of magnetization, field and torques.}
 \label{figS2}
\end{figure}

The magnetization dynamics can be described by the Landau-Lifshitz-Gilbert equation:
\begin{equation}
    {d\mathbf{m} \over dt} = -\gamma\mu_0 \mathbf{m}\times \mathbf{H_{eff}} + \Lambda \mathbf{m}\times\mathbf{\sigma}\times\mathbf{m}+\alpha \mathbf{m}\times{d\mathbf{m} \over dt}
\label{eq04}
\end{equation}
where $\mathbf{H_{eff}}=\mathbf{H_B}+\mathbf{h_{rf}}-\mu_0M_s\mathbf{m_z}$ is the total effective field, $\mathbf{h_{rf}}=(1/2)\mathbf{j_{rf}}d_{NM}\times\hat{y}$, $\mathbf{\sigma}=\theta_{SH}\mathbf{j_{rf}}\times\hat{y}$. Here $\mathbf{j_{rf}}$ denotes the microwave current density flowing through the bottom HM layer, and $\mathbf{\sigma}$ denotes the polarization of the spin current injected from HM into Py layer.

With the coordinate defined in Fig. \ref{figS2}, we can rewrite Eq. (\ref{eq04}) in the matrix form:
\begin{equation}
{d\over dt}
\begin{pmatrix}
m_y  \\
m_z
\end{pmatrix}
=-\gamma\mu_0H_B
\begin{pmatrix}
m_z \\
-m_y
\end{pmatrix}
+\gamma\mu_0M_s
\begin{pmatrix}
-m_z \\
0
\end{pmatrix}
+\gamma\mu_0{j_{rf} \over 2}
\begin{pmatrix}
\Lambda\theta_{SH} \\
d_{NM}
\end{pmatrix}
+\alpha{d\over dt}
\begin{pmatrix}
-m_z \\
m_y
\end{pmatrix}
\label{eq05}
\end{equation}
If we take $dm_y/dt = i\omega m_y$, $dm_z/dt = i\omega m_z$ and assuming $\alpha \ll 1$, from Eq. (\ref{eq05}) we can obtain:

\begin{subequations}
\begin{align}
    m_y &= \gamma\mu_0{j_{rf} \over 2} {\Lambda\theta_{SH}i\omega -(\omega_H+\omega_M)d_{NM} \over \omega_H(\omega_H+\omega_M)-\omega^2+i\alpha\omega(2\omega_H+\omega_M)} \\
    m_z &= \gamma\mu_0{j_{rf} \over 2} {i\omega d_{NM}+\Lambda\theta_{SH}\omega_H \over \omega_H(\omega_H+\omega_M)-\omega^2+i\alpha\omega(2\omega_H+\omega_M)}
\end{align}
\label{eq06}
\end{subequations}
where $\omega_M=\gamma\mu_0M_s$, $\omega_H=\gamma\mu_0H_B$. Near FMR, we have the Kittel equation $\omega^2 = \omega_H(\omega_H+\omega_M)$ and Eq. (\ref{eq06}a) can be approximated as:
\begin{equation}
{m_y} \approx i\gamma\mu_0{j_{rf}\over 2}\cdot{(\omega_H+\omega_M) \over (2\omega_H+\omega_M)\omega}\cdot {i\omega d_{NM}+\omega_H (\Lambda\theta_{SH}) \over (H_B - H_{res})+i\alpha\omega/\gamma}
\label{eq07}
\end{equation}
Eq. (\ref{eq07}) recovers Eq. (\ref{eq02}) in the main text. \\

\textbf{3. $V_O$ and $V_E$:}

As shown in Ref. \cite{YoonPRB2016}, the optical signal $V_O$ is proportional to the product of the polar magnetization ($m_y$) and the laser intensity $I$. The laser intensity is modulated at the microwave frequency $\omega$ as $\tilde{I} = I_0 (1+e^{i(\omega t+\phi_L)})/2$ where $\phi_L$ comes from the phase accumulation from the optical path and is a constant throughout the measurements. The polar magnetization can be similarly expressed in a complex way as $\tilde{m}_y=m^0_ye^{i(\omega t+\phi_M)}$ where $\phi_M$ is the precessional phase of the magnetization that we are interesed in. Thus we have:
\begin{equation}
V_O \sim \operatorname{Re}\left[ \tilde{I}\cdot \tilde{m}^*_y \right] \sim \operatorname{Re}[e^{-i(\phi_M-\phi_L)}]
\label{eq08}
\end{equation}
The electrical signal $V_E$ mainly comes from the anisotropic magnetoresistance ($R_A$) modulated by the microwave current $\tilde{j}^F_{rf}=j^F_{rf}e^{i\phi^F_j}$ flowing through the ferromagnet. $R_A$, which comes from varying $m_z$, can be treated as proportional to $m_y$ with an additional constant phase of $\pi/2$.
\begin{equation}
V_E \sim \operatorname{Re}\left[ \tilde{j}^F_{rf}\cdot \tilde{m}^*_y \right] \sim \operatorname{Re}[e^{-i(\phi_M-\phi^F_j)}]
\label{eq09}
\end{equation}
For both cases, when the biasing field goes from below to above resonance, an additional phase shift of $-\pi$ is added to $\phi_M$, which is reflected from the denominator in Eq. (\ref{eq07}).

The main difference from Eq. (\ref{eq08}) is that $\phi^F_j$ may not be a constant in the microwave circuit, such as in the two-bar configuration (Fig. 4) or in a complex multilayer structure. In addition, shunting effect from more conductive layers such as in Pt/Cu/Py (Fig. 5c) may significantly reduce the sensitivity of $\tilde{j}^F_{rf}$.\\

\textbf{4. Modulation harmonics and the lock-in \textit{X} and \textit{Y}:}

\begin{figure}[htb]
 \centering
 \includegraphics[width=6.0 in]{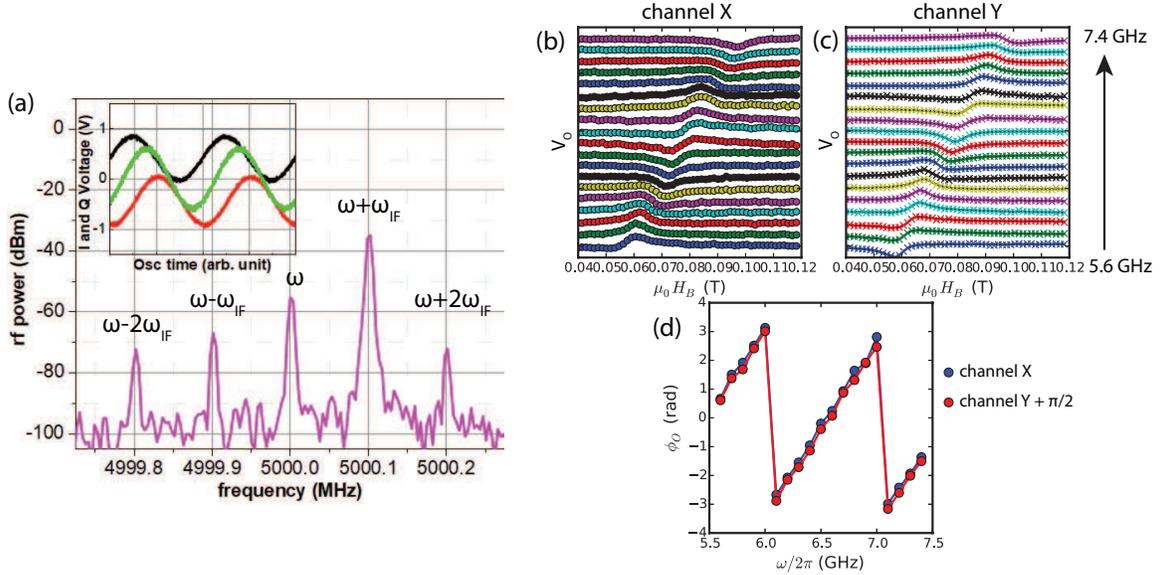}
 \caption{(a) Optimized microwave-power spectra after the IQ-mixing for $\omega/2\pi=5$ GHz and $\omega_{IF}/2\pi=100$ kHz, achieved by the appropriate I and Q parameters described in Section 1 and also simultaneously monitored from an oscilloscope (inset). (b-c) Optical signals from channel $X$ and $Y$ measured by a lock-in amplifier. (d) Extracted phase $\phi_O$ from (b) and (c).}
 \label{figS3}
\end{figure}

We confirm the negligible role of the IQ-mixing induced harmonics in the FMR excitation and the optical detection. The low-frequency modulation ($\omega_{IF}/2\pi=100$ kHz) induced from the IQ-mixing process introduces harmonic sidebands, which are shown in Fig.\ref{figS3}. For our heterodyne measurement, we excite FMR by using the upper sideband, i.e. $\omega$ + $\omega_{IF}$. This is achieved by fine-tuning the I and Q parameters (amplitude, offset, and phase) in the arbitrary function generator, see inset Fig. \ref{figS3}(a).

As an example, Fig. \ref{figS3}(a) shows the harmonics measured at $\omega/2\pi=5$ GHz by using a real-time spectrum analyzer. A clear dominance of the upper sideband $\omega+\omega_{IF}$ is achieved, more than 20 dB stronger than the other harmonic signals, such as the central-band, $\omega$, and lower-band, $\omega$ - $\omega_{IF}$. Other higher-order harmonics are even weaker. Nevertheless, we still examine any possible effects that could be induced by these harmonic signals. We compare the extracted phases from the optical detection lock-in channels $X$ and $Y$ (Figs. \ref{figS3}, b-c). The two phases are accurately separated by $\pi/2$ as shown in Fig. \ref{figS3}(d), which confirms the dominance of the single, upper-sideband excitation for the heterodyne measurements.

\textbf{5. Simulation of the 1-bar and 2-bar devices:}

\begin{figure}[htb]
 \centering
 \includegraphics[width=4.5 in]{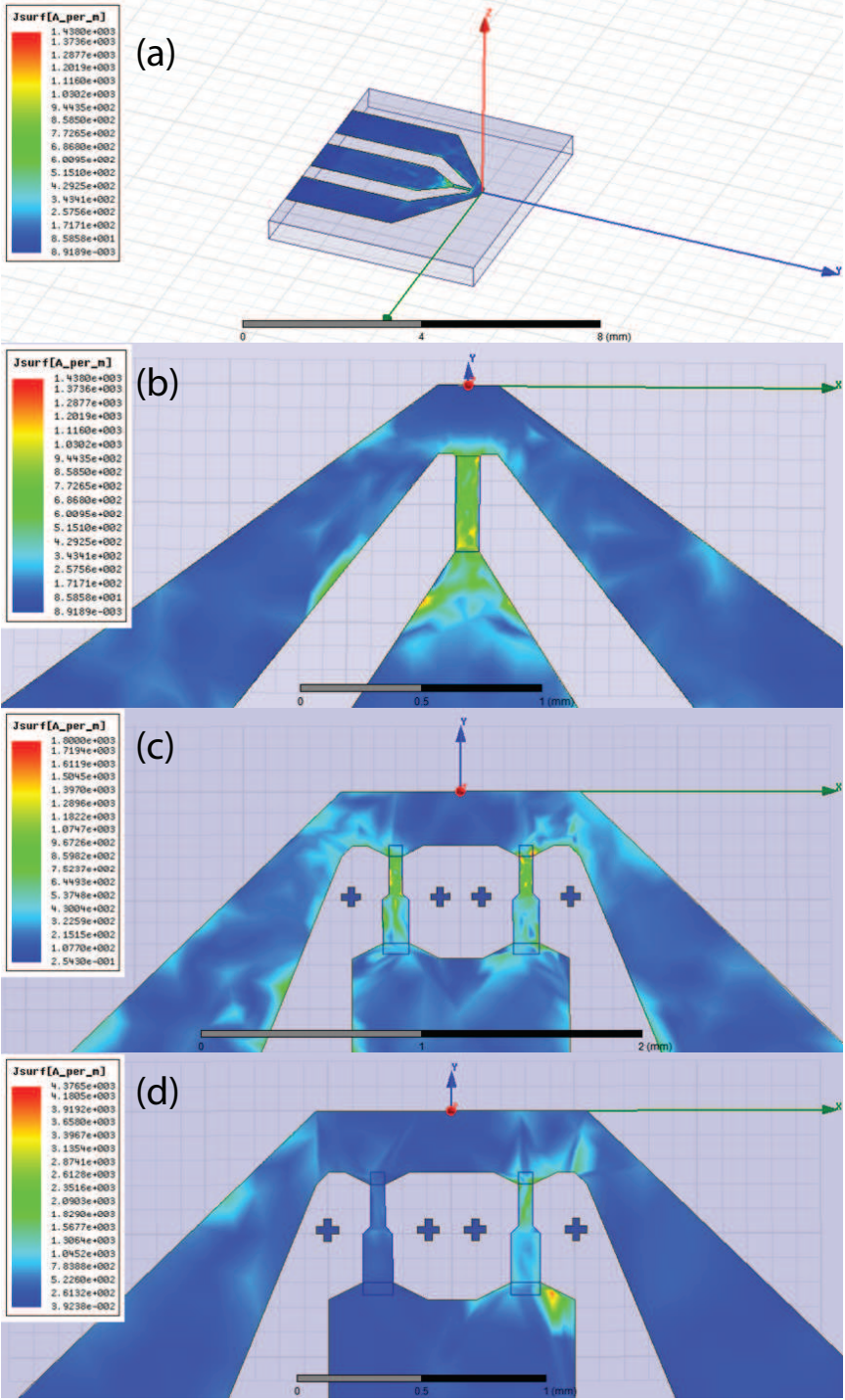}
 \caption{HFSS simulations. See the text for details.}
 \label{figS4}
\end{figure}

In order to examine the microwave field distribution and to check whether there is any local phase variation of the RF field induced by the electrical circuit, we have also conducted electromagnetic simulations using the 3D Electromagnetic Field High Frequency Structure Simulator (HFSS). Fig. \ref{figS4}(a) shows the simulated geometry which corresponds to our experimental materials and devices. The shorted coplanar waveguide and the center sample stripe are placed on a 300-nm SiO$_2$ ($\epsilon_r=4$) on a 500-$\mu$m Si ($\epsilon_r=11.9$) substrate. The resistivity of the CPW sheet is 2000 S/m, and the resistance of the Cu/Py and Pt/Py bars are set to be 26.9 $\Omega$ and 102.6 $\Omega$, respectively. Fig. \ref{figS4}(b) shows the RF surface current density ($J_{surf}$) of a single Pt/Py bar device. Strong density of $J_{surf}$ are symmetrically distributed at the device bar, which provides the driving microwave field for the magnetization dynamics. Fig. \ref{figS4}(c) shows the RF surface current density ($J_{surf}$) for two parallel Pt/Py bars. The $J_{surf}$ distributions on the two bars are the same in terms of both intensities and phases (see animation). Fig. \ref{figS4}(d) shows the RF surface current density ($J_{surf}$) for one Pt/Py and one Cu/Py bars which have very different resistance (Cu/Py on the right and Pt/Py on the left). Despite the large impedance variation along the two bars and asymmetric distributions of the $J_{surf}$, negligible phase lags are found throughout the microwave circuits, which means there is no additional phase offset from the circuit geometry in our experimental studies.

\begin{thebibliography}{68}%
\makeatletter
\providecommand \@ifxundefined [1]{%
 \@ifx{#1\undefined}
}%
\providecommand \@ifnum [1]{%
 \ifnum #1\expandafter \@firstoftwo
 \else \expandafter \@secondoftwo
 \fi
}%
\providecommand \@ifx [1]{%
 \ifx #1\expandafter \@firstoftwo
 \else \expandafter \@secondoftwo
 \fi
}%
\providecommand \natexlab [1]{#1}%
\providecommand \enquote  [1]{``#1''}%
\providecommand \bibnamefont  [1]{#1}%
\providecommand \bibfnamefont [1]{#1}%
\providecommand \citenamefont [1]{#1}%
\providecommand \href@noop [0]{\@secondoftwo}%
\providecommand \href [0]{\begingroup \@sanitize@url \@href}%
\providecommand \@href[1]{\@@startlink{#1}\@@href}%
\providecommand \@@href[1]{\endgroup#1\@@endlink}%
\providecommand \@sanitize@url [0]{\catcode `\\12\catcode `\$12\catcode
  `\&12\catcode `\#12\catcode `\^12\catcode `\_12\catcode `\%12\relax}%
\providecommand \@@startlink[1]{}%
\providecommand \@@endlink[0]{}%
\providecommand \url  [0]{\begingroup\@sanitize@url \@url }%
\providecommand \@url [1]{\endgroup\@href {#1}{\urlprefix }}%
\providecommand \urlprefix  [0]{URL }%
\providecommand \Eprint [0]{\href }%
\providecommand \doibase [0]{http://dx.doi.org/}%
\providecommand \selectlanguage [0]{\@gobble}%
\providecommand \bibinfo  [0]{\@secondoftwo}%
\providecommand \bibfield  [0]{\@secondoftwo}%
\providecommand \translation [1]{[#1]}%
\providecommand \BibitemOpen [0]{}%
\providecommand \bibitemStop [0]{}%
\providecommand \bibitemNoStop [0]{.\EOS\space}%
\providecommand \EOS [0]{\spacefactor3000\relax}%
\providecommand \BibitemShut  [1]{\csname bibitem#1\endcsname}%
\let\auto@bib@innerbib\@empty
\bibitem [{\citenamefont {Hoffmann}\ and\ \citenamefont
  {Bader}(2015)}]{HoffmannPRApplied2015}%
  \BibitemOpen
  \bibfield  {author} {\bibinfo {author} {\bibfnamefont {A.}~\bibnamefont
  {Hoffmann}}\ and\ \bibinfo {author} {\bibfnamefont {S.~D.}\ \bibnamefont
  {Bader}},\ }\href {\doibase https://doi.org/10.1103/PhysRevApplied.4.047001}
  {\bibfield  {journal} {\bibinfo  {journal} {Phys. Rev. Applied}\ }\textbf
  {\bibinfo {volume} {4}},\ \bibinfo {pages} {047001} (\bibinfo {year}
  {2015})}\BibitemShut {NoStop}%
\bibitem [{\citenamefont {Cornelissen}\ \emph {et~al.}(2015)\citenamefont
  {Cornelissen}, \citenamefont {Liu}, \citenamefont {Duine}, \citenamefont
  {Ben~Youssef},\ and\ \citenamefont {van Wees}}]{CornelissenNPhys2015}%
  \BibitemOpen
  \bibfield  {author} {\bibinfo {author} {\bibfnamefont {L.~J.}\ \bibnamefont
  {Cornelissen}}, \bibinfo {author} {\bibfnamefont {J.}~\bibnamefont {Liu}},
  \bibinfo {author} {\bibfnamefont {R.~A.}\ \bibnamefont {Duine}}, \bibinfo
  {author} {\bibfnamefont {J.}~\bibnamefont {Ben~Youssef}}, \ and\ \bibinfo
  {author} {\bibfnamefont {B.~J.}\ \bibnamefont {van Wees}},\ }\href@noop {}
  {\bibfield  {journal} {\bibinfo  {journal} {Nature Physics}\ }\textbf
  {\bibinfo {volume} {11}},\ \bibinfo {pages} {1022–1026} (\bibinfo {year}
  {2015})}\BibitemShut {NoStop}%
\bibitem [{\citenamefont {Wesenberg}\ \emph {et~al.}(2017)\citenamefont
  {Wesenberg}, \citenamefont {Liu}, \citenamefont {Balzar}, \citenamefont
  {Wu},\ and\ \citenamefont {Zink}}]{WesenbergNPhys2017}%
  \BibitemOpen
  \bibfield  {author} {\bibinfo {author} {\bibfnamefont {D.}~\bibnamefont
  {Wesenberg}}, \bibinfo {author} {\bibfnamefont {T.}~\bibnamefont {Liu}},
  \bibinfo {author} {\bibfnamefont {D.}~\bibnamefont {Balzar}}, \bibinfo
  {author} {\bibfnamefont {M.}~\bibnamefont {Wu}}, \ and\ \bibinfo {author}
  {\bibfnamefont {B.~L.}\ \bibnamefont {Zink}},\ }\href@noop {} {\bibfield
  {journal} {\bibinfo  {journal} {Nature Physics}\ }\textbf {\bibinfo {volume}
  {13}},\ \bibinfo {pages} {987} (\bibinfo {year} {2017})}\BibitemShut
  {NoStop}%
\bibitem [{\citenamefont {Lebrun}\ \emph {et~al.}(2018)\citenamefont {Lebrun},
  \citenamefont {Ross}, \citenamefont {Bender}, \citenamefont {Qaiumzadeh},
  \citenamefont {Baldrati}, \citenamefont {Cramer}, \citenamefont {Brataas},
  \citenamefont {Duine},\ and\ \citenamefont {Kl\:{a}ui}}]{LebrunNature2018}%
  \BibitemOpen
  \bibfield  {author} {\bibinfo {author} {\bibfnamefont {R.}~\bibnamefont
  {Lebrun}}, \bibinfo {author} {\bibfnamefont {A.}~\bibnamefont {Ross}},
  \bibinfo {author} {\bibfnamefont {S.~A.}\ \bibnamefont {Bender}}, \bibinfo
  {author} {\bibfnamefont {A.}~\bibnamefont {Qaiumzadeh}}, \bibinfo {author}
  {\bibfnamefont {L.}~\bibnamefont {Baldrati}}, \bibinfo {author}
  {\bibfnamefont {J.}~\bibnamefont {Cramer}}, \bibinfo {author} {\bibfnamefont
  {A.}~\bibnamefont {Brataas}}, \bibinfo {author} {\bibfnamefont {R.~A.}\
  \bibnamefont {Duine}}, \ and\ \bibinfo {author} {\bibfnamefont
  {M.}~\bibnamefont {Kl\:{a}ui}},\ }\href@noop {} {\bibfield  {journal}
  {\bibinfo  {journal} {Nature}\ }\textbf {\bibinfo {volume} {561}},\ \bibinfo
  {pages} {222} (\bibinfo {year} {2018})}\BibitemShut {NoStop}%
\bibitem [{\citenamefont {Kruglyak}\ \emph {et~al.}(2010)\citenamefont
  {Kruglyak}, \citenamefont {Demokritov},\ and\ \citenamefont
  {Grundler}}]{KruglyakJPD2010}%
  \BibitemOpen
  \bibfield  {author} {\bibinfo {author} {\bibfnamefont {V.~V.}\ \bibnamefont
  {Kruglyak}}, \bibinfo {author} {\bibfnamefont {S.~O.}\ \bibnamefont
  {Demokritov}}, \ and\ \bibinfo {author} {\bibfnamefont {D.}~\bibnamefont
  {Grundler}},\ }\href {http://stacks.iop.org/0022-3727/43/i=26/a=264001}
  {\bibfield  {journal} {\bibinfo  {journal} {J Phys. D: Appl. Phys.}\ }\textbf
  {\bibinfo {volume} {43}},\ \bibinfo {pages} {264001} (\bibinfo {year}
  {2010})}\BibitemShut {NoStop}%
\bibitem [{\citenamefont {Perzlmaier}\ \emph {et~al.}(2008)\citenamefont
  {Perzlmaier}, \citenamefont {Woltersdorf},\ and\ \citenamefont
  {Back}}]{PerzlmaierPRB2008}%
  \BibitemOpen
  \bibfield  {author} {\bibinfo {author} {\bibfnamefont {K.}~\bibnamefont
  {Perzlmaier}}, \bibinfo {author} {\bibfnamefont {G.}~\bibnamefont
  {Woltersdorf}}, \ and\ \bibinfo {author} {\bibfnamefont {C.~H.}\ \bibnamefont
  {Back}},\ }\href {\doibase 10.1103/PhysRevB.77.054425} {\bibfield  {journal}
  {\bibinfo  {journal} {Phys. Rev. B}\ }\textbf {\bibinfo {volume} {77}},\
  \bibinfo {pages} {054425} (\bibinfo {year} {2008})}\BibitemShut {NoStop}%
\bibitem [{\citenamefont {Schneider}\ \emph {et~al.}(2008)\citenamefont
  {Schneider}, \citenamefont {Serga}, \citenamefont {Leven}, \citenamefont
  {Hillebrands}, \citenamefont {Stamps},\ and\ \citenamefont
  {Kostylev}}]{SchneiderAPL2008}%
  \BibitemOpen
  \bibfield  {author} {\bibinfo {author} {\bibfnamefont {T.}~\bibnamefont
  {Schneider}}, \bibinfo {author} {\bibfnamefont {A.~A.}\ \bibnamefont
  {Serga}}, \bibinfo {author} {\bibfnamefont {B.}~\bibnamefont {Leven}},
  \bibinfo {author} {\bibfnamefont {B.}~\bibnamefont {Hillebrands}}, \bibinfo
  {author} {\bibfnamefont {R.~L.}\ \bibnamefont {Stamps}}, \ and\ \bibinfo
  {author} {\bibfnamefont {M.~P.}\ \bibnamefont {Kostylev}},\ }\href {\doibase
  10.1063/1.2834714} {\bibfield  {journal} {\bibinfo  {journal} {Appl. Phys.
  Lett.}\ }\textbf {\bibinfo {volume} {92}},\ \bibinfo {pages} {022505}
  (\bibinfo {year} {2008})}\BibitemShut {NoStop}%
\bibitem [{\citenamefont {Kaka}\ \emph {et~al.}(2005)\citenamefont {Kaka},
  \citenamefont {Pufall}, \citenamefont {Rippard}, \citenamefont {Silva},
  \citenamefont {Russek},\ and\ \citenamefont {Katine}}]{KakaNature2005}%
  \BibitemOpen
  \bibfield  {author} {\bibinfo {author} {\bibfnamefont {S.}~\bibnamefont
  {Kaka}}, \bibinfo {author} {\bibfnamefont {M.~R.}\ \bibnamefont {Pufall}},
  \bibinfo {author} {\bibfnamefont {W.~H.}\ \bibnamefont {Rippard}}, \bibinfo
  {author} {\bibfnamefont {T.~J.}\ \bibnamefont {Silva}}, \bibinfo {author}
  {\bibfnamefont {S.~E.}\ \bibnamefont {Russek}}, \ and\ \bibinfo {author}
  {\bibfnamefont {J.~A.}\ \bibnamefont {Katine}},\ }\href@noop {} {\bibfield
  {journal} {\bibinfo  {journal} {Nature}\ }\textbf {\bibinfo {volume} {437}},\
  \bibinfo {pages} {389} (\bibinfo {year} {2005})}\BibitemShut {NoStop}%
\bibitem [{\citenamefont {Mancoff}\ \emph {et~al.}(2005)\citenamefont
  {Mancoff}, \citenamefont {Rizzo}, \citenamefont {Engel},\ and\ \citenamefont
  {Tehrani}}]{MancoffNature2005}%
  \BibitemOpen
  \bibfield  {author} {\bibinfo {author} {\bibfnamefont {F.~B.}\ \bibnamefont
  {Mancoff}}, \bibinfo {author} {\bibfnamefont {N.~D.}\ \bibnamefont {Rizzo}},
  \bibinfo {author} {\bibfnamefont {B.~N.}\ \bibnamefont {Engel}}, \ and\
  \bibinfo {author} {\bibfnamefont {S.}~\bibnamefont {Tehrani}},\ }\href@noop
  {} {\bibfield  {journal} {\bibinfo  {journal} {Nature}\ }\textbf {\bibinfo
  {volume} {437}},\ \bibinfo {pages} {393} (\bibinfo {year}
  {2005})}\BibitemShut {NoStop}%
\bibitem [{\citenamefont {Li}\ \emph {et~al.}(2017)\citenamefont {Li},
  \citenamefont {de~Milly}, \citenamefont {Abreu~Araujo}, \citenamefont
  {Klein}, \citenamefont {Cros}, \citenamefont {Grollier},\ and\ \citenamefont
  {de~Loubens}}]{LiPRL2017}%
  \BibitemOpen
  \bibfield  {author} {\bibinfo {author} {\bibfnamefont {Y.}~\bibnamefont
  {Li}}, \bibinfo {author} {\bibfnamefont {X.}~\bibnamefont {de~Milly}},
  \bibinfo {author} {\bibfnamefont {F.}~\bibnamefont {Abreu~Araujo}}, \bibinfo
  {author} {\bibfnamefont {O.}~\bibnamefont {Klein}}, \bibinfo {author}
  {\bibfnamefont {V.}~\bibnamefont {Cros}}, \bibinfo {author} {\bibfnamefont
  {J.}~\bibnamefont {Grollier}}, \ and\ \bibinfo {author} {\bibfnamefont
  {G.}~\bibnamefont {de~Loubens}},\ }\href {\doibase
  10.1103/PhysRevLett.118.247202} {\bibfield  {journal} {\bibinfo  {journal}
  {Phys. Rev. Lett.}\ }\textbf {\bibinfo {volume} {118}},\ \bibinfo {pages}
  {247202} (\bibinfo {year} {2017})}\BibitemShut {NoStop}%
\bibitem [{\citenamefont {Klingler}\ \emph {et~al.}(2018)\citenamefont
  {Klingler}, \citenamefont {Amin}, \citenamefont {Gepr\"ags}, \citenamefont
  {Ganzhorn}, \citenamefont {Maier-Flaig}, \citenamefont {Althammer},
  \citenamefont {Huebl}, \citenamefont {Gross}, \citenamefont {McMichael},
  \citenamefont {Stiles}, \citenamefont {Goennenwein},\ and\ \citenamefont
  {Weiler}}]{KlinglerPRL2018}%
  \BibitemOpen
  \bibfield  {author} {\bibinfo {author} {\bibfnamefont {S.}~\bibnamefont
  {Klingler}}, \bibinfo {author} {\bibfnamefont {V.}~\bibnamefont {Amin}},
  \bibinfo {author} {\bibfnamefont {S.}~\bibnamefont {Gepr\"ags}}, \bibinfo
  {author} {\bibfnamefont {K.}~\bibnamefont {Ganzhorn}}, \bibinfo {author}
  {\bibfnamefont {H.}~\bibnamefont {Maier-Flaig}}, \bibinfo {author}
  {\bibfnamefont {M.}~\bibnamefont {Althammer}}, \bibinfo {author}
  {\bibfnamefont {H.}~\bibnamefont {Huebl}}, \bibinfo {author} {\bibfnamefont
  {R.}~\bibnamefont {Gross}}, \bibinfo {author} {\bibfnamefont {R.~D.}\
  \bibnamefont {McMichael}}, \bibinfo {author} {\bibfnamefont {M.~D.}\
  \bibnamefont {Stiles}}, \bibinfo {author} {\bibfnamefont {S.~T.~B.}\
  \bibnamefont {Goennenwein}}, \ and\ \bibinfo {author} {\bibfnamefont
  {M.}~\bibnamefont {Weiler}},\ }\href {\doibase
  10.1103/PhysRevLett.120.127201} {\bibfield  {journal} {\bibinfo  {journal}
  {Phys. Rev. Lett.}\ }\textbf {\bibinfo {volume} {120}},\ \bibinfo {pages}
  {127201} (\bibinfo {year} {2018})}\BibitemShut {NoStop}%
\bibitem [{\citenamefont {Chen}\ \emph {et~al.}(2018)\citenamefont {Chen},
  \citenamefont {Liu}, \citenamefont {Liu}, \citenamefont {Xiao}, \citenamefont
  {Xia}, \citenamefont {Bauer}, \citenamefont {Wu},\ and\ \citenamefont
  {Yu}}]{ChenPRL2018}%
  \BibitemOpen
  \bibfield  {author} {\bibinfo {author} {\bibfnamefont {J.}~\bibnamefont
  {Chen}}, \bibinfo {author} {\bibfnamefont {C.}~\bibnamefont {Liu}}, \bibinfo
  {author} {\bibfnamefont {T.}~\bibnamefont {Liu}}, \bibinfo {author}
  {\bibfnamefont {Y.}~\bibnamefont {Xiao}}, \bibinfo {author} {\bibfnamefont
  {K.}~\bibnamefont {Xia}}, \bibinfo {author} {\bibfnamefont {G.~E.~W.}\
  \bibnamefont {Bauer}}, \bibinfo {author} {\bibfnamefont {M.}~\bibnamefont
  {Wu}}, \ and\ \bibinfo {author} {\bibfnamefont {H.}~\bibnamefont {Yu}},\
  }\href {\doibase 10.1103/PhysRevLett.120.217202} {\bibfield  {journal}
  {\bibinfo  {journal} {Phys. Rev. Lett.}\ }\textbf {\bibinfo {volume} {120}},\
  \bibinfo {pages} {217202} (\bibinfo {year} {2018})}\BibitemShut {NoStop}%
\bibitem [{\citenamefont {Huebl}\ \emph {et~al.}(2013)\citenamefont {Huebl},
  \citenamefont {Zollitsch}, \citenamefont {Lotze}, \citenamefont {Hocke},
  \citenamefont {Greifenstein}, \citenamefont {Marx}, \citenamefont {Gross},\
  and\ \citenamefont {Goennenwein}}]{HueblPRL2013}%
  \BibitemOpen
  \bibfield  {author} {\bibinfo {author} {\bibfnamefont {H.}~\bibnamefont
  {Huebl}}, \bibinfo {author} {\bibfnamefont {C.~W.}\ \bibnamefont
  {Zollitsch}}, \bibinfo {author} {\bibfnamefont {J.}~\bibnamefont {Lotze}},
  \bibinfo {author} {\bibfnamefont {F.}~\bibnamefont {Hocke}}, \bibinfo
  {author} {\bibfnamefont {M.}~\bibnamefont {Greifenstein}}, \bibinfo {author}
  {\bibfnamefont {A.}~\bibnamefont {Marx}}, \bibinfo {author} {\bibfnamefont
  {R.}~\bibnamefont {Gross}}, \ and\ \bibinfo {author} {\bibfnamefont
  {S.~T.~B.}\ \bibnamefont {Goennenwein}},\ }\href {\doibase
  10.1103/PhysRevLett.111.127003} {\bibfield  {journal} {\bibinfo  {journal}
  {Phys. Rev. Lett.}\ }\textbf {\bibinfo {volume} {111}},\ \bibinfo {pages}
  {127003} (\bibinfo {year} {2013})}\BibitemShut {NoStop}%
\bibitem [{\citenamefont {Tabuchi}\ \emph {et~al.}(2014)\citenamefont
  {Tabuchi}, \citenamefont {Ishino}, \citenamefont {Ishikawa}, \citenamefont
  {Yamazaki}, \citenamefont {Usami},\ and\ \citenamefont
  {Nakamura}}]{TabuchiPRL2014}%
  \BibitemOpen
  \bibfield  {author} {\bibinfo {author} {\bibfnamefont {Y.}~\bibnamefont
  {Tabuchi}}, \bibinfo {author} {\bibfnamefont {S.}~\bibnamefont {Ishino}},
  \bibinfo {author} {\bibfnamefont {T.}~\bibnamefont {Ishikawa}}, \bibinfo
  {author} {\bibfnamefont {R.}~\bibnamefont {Yamazaki}}, \bibinfo {author}
  {\bibfnamefont {K.}~\bibnamefont {Usami}}, \ and\ \bibinfo {author}
  {\bibfnamefont {Y.}~\bibnamefont {Nakamura}},\ }\href {\doibase
  10.1103/PhysRevLett.113.083603} {\bibfield  {journal} {\bibinfo  {journal}
  {Phys. Rev. Lett.}\ }\textbf {\bibinfo {volume} {113}},\ \bibinfo {pages}
  {083603} (\bibinfo {year} {2014})}\BibitemShut {NoStop}%
\bibitem [{\citenamefont {Zhang}\ \emph {et~al.}(2014)\citenamefont {Zhang},
  \citenamefont {Zou}, \citenamefont {Jiang},\ and\ \citenamefont
  {Tang}}]{ZhangPRL2014}%
  \BibitemOpen
  \bibfield  {author} {\bibinfo {author} {\bibfnamefont {X.}~\bibnamefont
  {Zhang}}, \bibinfo {author} {\bibfnamefont {C.-L.}\ \bibnamefont {Zou}},
  \bibinfo {author} {\bibfnamefont {L.}~\bibnamefont {Jiang}}, \ and\ \bibinfo
  {author} {\bibfnamefont {H.~X.}\ \bibnamefont {Tang}},\ }\href {\doibase
  10.1103/PhysRevLett.113.156401} {\bibfield  {journal} {\bibinfo  {journal}
  {Phys. Rev. Lett.}\ }\textbf {\bibinfo {volume} {113}},\ \bibinfo {pages}
  {156401} (\bibinfo {year} {2014})}\BibitemShut {NoStop}%
\bibitem [{\citenamefont {Bai}\ \emph {et~al.}(2015)\citenamefont {Bai},
  \citenamefont {Harder}, \citenamefont {Chen}, \citenamefont {Fan},
  \citenamefont {Xiao},\ and\ \citenamefont {Hu}}]{BaiPRL2015}%
  \BibitemOpen
  \bibfield  {author} {\bibinfo {author} {\bibfnamefont {L.}~\bibnamefont
  {Bai}}, \bibinfo {author} {\bibfnamefont {M.}~\bibnamefont {Harder}},
  \bibinfo {author} {\bibfnamefont {Y.~P.}\ \bibnamefont {Chen}}, \bibinfo
  {author} {\bibfnamefont {X.}~\bibnamefont {Fan}}, \bibinfo {author}
  {\bibfnamefont {J.~Q.}\ \bibnamefont {Xiao}}, \ and\ \bibinfo {author}
  {\bibfnamefont {C.-M.}\ \bibnamefont {Hu}},\ }\href {\doibase
  10.1103/PhysRevLett.114.227201} {\bibfield  {journal} {\bibinfo  {journal}
  {Phys. Rev. Lett.}\ }\textbf {\bibinfo {volume} {114}},\ \bibinfo {pages}
  {227201} (\bibinfo {year} {2015})}\BibitemShut {NoStop}%
\bibitem [{\citenamefont {Zhang}\ \emph {et~al.}(2016)\citenamefont {Zhang},
  \citenamefont {Zou}, \citenamefont {Jiang},\ and\ \citenamefont
  {Tang}}]{ZhangScienceAdv2016}%
  \BibitemOpen
  \bibfield  {author} {\bibinfo {author} {\bibfnamefont {X.}~\bibnamefont
  {Zhang}}, \bibinfo {author} {\bibfnamefont {C.-L.}\ \bibnamefont {Zou}},
  \bibinfo {author} {\bibfnamefont {L.}~\bibnamefont {Jiang}}, \ and\ \bibinfo
  {author} {\bibfnamefont {H.~X.}\ \bibnamefont {Tang}},\ }\href {\doibase
  10.1126/sciadv.1501286} {\bibfield  {journal} {\bibinfo  {journal} {Science
  Advances}\ }\textbf {\bibinfo {volume} {2}} (\bibinfo {year} {2016}),\
  10.1126/sciadv.1501286}\BibitemShut {NoStop}%
\bibitem [{\citenamefont {Kikkawa}\ \emph {et~al.}(2016)\citenamefont
  {Kikkawa}, \citenamefont {Shen}, \citenamefont {Flebus}, \citenamefont
  {Duine}, \citenamefont {Uchida}, \citenamefont {Qiu}, \citenamefont {Bauer},\
  and\ \citenamefont {Saitoh}}]{KikkawaPRL2016}%
  \BibitemOpen
  \bibfield  {author} {\bibinfo {author} {\bibfnamefont {T.}~\bibnamefont
  {Kikkawa}}, \bibinfo {author} {\bibfnamefont {K.}~\bibnamefont {Shen}},
  \bibinfo {author} {\bibfnamefont {B.}~\bibnamefont {Flebus}}, \bibinfo
  {author} {\bibfnamefont {R.~A.}\ \bibnamefont {Duine}}, \bibinfo {author}
  {\bibfnamefont {K.-i.}\ \bibnamefont {Uchida}}, \bibinfo {author}
  {\bibfnamefont {Z.}~\bibnamefont {Qiu}}, \bibinfo {author} {\bibfnamefont
  {G.~E.~W.}\ \bibnamefont {Bauer}}, \ and\ \bibinfo {author} {\bibfnamefont
  {E.}~\bibnamefont {Saitoh}},\ }\href {\doibase
  10.1103/PhysRevLett.117.207203} {\bibfield  {journal} {\bibinfo  {journal}
  {Phys. Rev. Lett.}\ }\textbf {\bibinfo {volume} {117}},\ \bibinfo {pages}
  {207203} (\bibinfo {year} {2016})}\BibitemShut {NoStop}%
\bibitem [{\citenamefont {Sklenar}\ \emph {et~al.}(2018)\citenamefont
  {Sklenar}, \citenamefont {Zhang}, \citenamefont {Jungfleisch},\ and\
  \citenamefont {Hoffmann}}]{SklenarSpintronicsXI2018}%
  \BibitemOpen
  \bibfield  {author} {\bibinfo {author} {\bibfnamefont {J.}~\bibnamefont
  {Sklenar}}, \bibinfo {author} {\bibfnamefont {W.}~\bibnamefont {Zhang}},
  \bibinfo {author} {\bibfnamefont {M.~B.}\ \bibnamefont {Jungfleisch}}, \ and\
  \bibinfo {author} {\bibfnamefont {A.}~\bibnamefont {Hoffmann}},\ }\href@noop
  {} {\bibfield  {journal} {\bibinfo  {journal} {Proc. SPIE}\ }\textbf
  {\bibinfo {volume} {10732}},\ \bibinfo {pages} {12} (\bibinfo {year}
  {2018})}\BibitemShut {NoStop}%
\bibitem [{\citenamefont {Sklenar}\ \emph {et~al.}(2017)\citenamefont
  {Sklenar}, \citenamefont {Zhang}, \citenamefont {Jungfleisch}, \citenamefont
  {Saglam}, \citenamefont {Grudichak}, \citenamefont {Jiang}, \citenamefont
  {Pearson}, \citenamefont {Ketterson},\ and\ \citenamefont
  {Hoffmann}}]{SklenarPRB2017Unidirectional}%
  \BibitemOpen
  \bibfield  {author} {\bibinfo {author} {\bibfnamefont {J.}~\bibnamefont
  {Sklenar}}, \bibinfo {author} {\bibfnamefont {W.}~\bibnamefont {Zhang}},
  \bibinfo {author} {\bibfnamefont {M.~B.}\ \bibnamefont {Jungfleisch}},
  \bibinfo {author} {\bibfnamefont {H.}~\bibnamefont {Saglam}}, \bibinfo
  {author} {\bibfnamefont {S.}~\bibnamefont {Grudichak}}, \bibinfo {author}
  {\bibfnamefont {W.}~\bibnamefont {Jiang}}, \bibinfo {author} {\bibfnamefont
  {J.~E.}\ \bibnamefont {Pearson}}, \bibinfo {author} {\bibfnamefont {J.~B.}\
  \bibnamefont {Ketterson}}, \ and\ \bibinfo {author} {\bibfnamefont
  {A.}~\bibnamefont {Hoffmann}},\ }\href@noop {} {\bibfield  {journal}
  {\bibinfo  {journal} {Phys. Rev. B}\ }\textbf {\bibinfo {volume} {95}},\
  \bibinfo {pages} {224431} (\bibinfo {year} {2017})}\BibitemShut {NoStop}%
\bibitem [{\citenamefont {Sinova}\ \emph {et~al.}(2015)\citenamefont {Sinova},
  \citenamefont {Valenzuela}, \citenamefont {Wunderlich}, \citenamefont
  {Back},\ and\ \citenamefont {Jungwirth}}]{SinovaRMP2015}%
  \BibitemOpen
  \bibfield  {author} {\bibinfo {author} {\bibfnamefont {J.}~\bibnamefont
  {Sinova}}, \bibinfo {author} {\bibfnamefont {S.~O.}\ \bibnamefont
  {Valenzuela}}, \bibinfo {author} {\bibfnamefont {J.}~\bibnamefont
  {Wunderlich}}, \bibinfo {author} {\bibfnamefont {C.~H.}\ \bibnamefont
  {Back}}, \ and\ \bibinfo {author} {\bibfnamefont {T.}~\bibnamefont
  {Jungwirth}},\ }\href {\doibase 10.1103/RevModPhys.87.1213} {\bibfield
  {journal} {\bibinfo  {journal} {Rev. Mod. Phys.}\ }\textbf {\bibinfo {volume}
  {87}},\ \bibinfo {pages} {1213} (\bibinfo {year} {2015})}\BibitemShut
  {NoStop}%
\bibitem [{\citenamefont {Sankey}\ \emph {et~al.}(2006)\citenamefont {Sankey},
  \citenamefont {Braganca}, \citenamefont {Garcia}, \citenamefont {Krivorotov},
  \citenamefont {Buhrman},\ and\ \citenamefont {Ralph}}]{SankeyPRL2006}%
  \BibitemOpen
  \bibfield  {author} {\bibinfo {author} {\bibfnamefont {J.~C.}\ \bibnamefont
  {Sankey}}, \bibinfo {author} {\bibfnamefont {P.~M.}\ \bibnamefont
  {Braganca}}, \bibinfo {author} {\bibfnamefont {A.~G.~F.}\ \bibnamefont
  {Garcia}}, \bibinfo {author} {\bibfnamefont {I.~N.}\ \bibnamefont
  {Krivorotov}}, \bibinfo {author} {\bibfnamefont {R.~A.}\ \bibnamefont
  {Buhrman}}, \ and\ \bibinfo {author} {\bibfnamefont {D.~C.}\ \bibnamefont
  {Ralph}},\ }\href {\doibase 10.1103/PhysRevLett.96.227601} {\bibfield
  {journal} {\bibinfo  {journal} {Phys. Rev. Lett.}\ }\textbf {\bibinfo
  {volume} {96}},\ \bibinfo {pages} {227601} (\bibinfo {year}
  {2006})}\BibitemShut {NoStop}%
\bibitem [{\citenamefont {Mosendz}\ \emph
  {et~al.}(2010{\natexlab{a}})\citenamefont {Mosendz}, \citenamefont {Pearson},
  \citenamefont {Fradin}, \citenamefont {Bauer}, \citenamefont {Bader},\ and\
  \citenamefont {Hoffmann}}]{MosendzPRL2010}%
  \BibitemOpen
  \bibfield  {author} {\bibinfo {author} {\bibfnamefont {O.}~\bibnamefont
  {Mosendz}}, \bibinfo {author} {\bibfnamefont {J.~E.}\ \bibnamefont
  {Pearson}}, \bibinfo {author} {\bibfnamefont {F.~Y.}\ \bibnamefont {Fradin}},
  \bibinfo {author} {\bibfnamefont {G.~E.~W.}\ \bibnamefont {Bauer}}, \bibinfo
  {author} {\bibfnamefont {S.~D.}\ \bibnamefont {Bader}}, \ and\ \bibinfo
  {author} {\bibfnamefont {A.}~\bibnamefont {Hoffmann}},\ }\href {\doibase
  10.1103/PhysRevLett.104.046601} {\bibfield  {journal} {\bibinfo  {journal}
  {Phys. Rev. Lett.}\ }\textbf {\bibinfo {volume} {104}},\ \bibinfo {pages}
  {046601} (\bibinfo {year} {2010}{\natexlab{a}})}\BibitemShut {NoStop}%
\bibitem [{\citenamefont {Liu}\ \emph {et~al.}(2011)\citenamefont {Liu},
  \citenamefont {Moriyama}, \citenamefont {Ralph},\ and\ \citenamefont
  {Buhrman}}]{LiuPRL2011}%
  \BibitemOpen
  \bibfield  {author} {\bibinfo {author} {\bibfnamefont {L.}~\bibnamefont
  {Liu}}, \bibinfo {author} {\bibfnamefont {T.}~\bibnamefont {Moriyama}},
  \bibinfo {author} {\bibfnamefont {D.~C.}\ \bibnamefont {Ralph}}, \ and\
  \bibinfo {author} {\bibfnamefont {R.~A.}\ \bibnamefont {Buhrman}},\ }\href
  {\doibase 10.1103/PhysRevLett.106.036601} {\bibfield  {journal} {\bibinfo
  {journal} {Phys. Rev. Lett.}\ }\textbf {\bibinfo {volume} {106}},\ \bibinfo
  {pages} {036601} (\bibinfo {year} {2011})}\BibitemShut {NoStop}%
\bibitem [{\citenamefont {Weiler}\ \emph {et~al.}(2014)\citenamefont {Weiler},
  \citenamefont {Shaw}, \citenamefont {Nembach},\ and\ \citenamefont
  {Silva}}]{WeilerPRL2014}%
  \BibitemOpen
  \bibfield  {author} {\bibinfo {author} {\bibfnamefont {M.}~\bibnamefont
  {Weiler}}, \bibinfo {author} {\bibfnamefont {J.~M.}\ \bibnamefont {Shaw}},
  \bibinfo {author} {\bibfnamefont {H.~T.}\ \bibnamefont {Nembach}}, \ and\
  \bibinfo {author} {\bibfnamefont {T.~J.}\ \bibnamefont {Silva}},\ }\href
  {\doibase 10.1103/PhysRevLett.113.157204} {\bibfield  {journal} {\bibinfo
  {journal} {Phys. Rev. Lett.}\ }\textbf {\bibinfo {volume} {113}},\ \bibinfo
  {pages} {157204} (\bibinfo {year} {2014})}\BibitemShut {NoStop}%
\bibitem [{\citenamefont {Bai}\ \emph {et~al.}(2013)\citenamefont {Bai},
  \citenamefont {Hyde}, \citenamefont {Gui}, \citenamefont {Hu}, \citenamefont
  {Vlaminck}, \citenamefont {Pearson}, \citenamefont {Bader},\ and\
  \citenamefont {Hoffmann}}]{BaiLHPRL2013}%
  \BibitemOpen
  \bibfield  {author} {\bibinfo {author} {\bibfnamefont {L.}~\bibnamefont
  {Bai}}, \bibinfo {author} {\bibfnamefont {P.}~\bibnamefont {Hyde}}, \bibinfo
  {author} {\bibfnamefont {Y.~S.}\ \bibnamefont {Gui}}, \bibinfo {author}
  {\bibfnamefont {C.-M.}\ \bibnamefont {Hu}}, \bibinfo {author} {\bibfnamefont
  {V.}~\bibnamefont {Vlaminck}}, \bibinfo {author} {\bibfnamefont {J.~E.}\
  \bibnamefont {Pearson}}, \bibinfo {author} {\bibfnamefont {S.~D.}\
  \bibnamefont {Bader}}, \ and\ \bibinfo {author} {\bibfnamefont
  {A.}~\bibnamefont {Hoffmann}},\ }\href@noop {} {\bibfield  {journal}
  {\bibinfo  {journal} {Phys. Rev. Lett.}\ }\textbf {\bibinfo {volume} {111}},\
  \bibinfo {pages} {217602} (\bibinfo {year} {2013})}\BibitemShut {NoStop}%
\bibitem [{\citenamefont {Harder}\ \emph {et~al.}(2011)\citenamefont {Harder},
  \citenamefont {Cao}, \citenamefont {Gui}, \citenamefont {Fan},\ and\
  \citenamefont {Hu}}]{HarderPRB2011}%
  \BibitemOpen
  \bibfield  {author} {\bibinfo {author} {\bibfnamefont {M.}~\bibnamefont
  {Harder}}, \bibinfo {author} {\bibfnamefont {Z.~X.}\ \bibnamefont {Cao}},
  \bibinfo {author} {\bibfnamefont {Y.~S.}\ \bibnamefont {Gui}}, \bibinfo
  {author} {\bibfnamefont {X.~L.}\ \bibnamefont {Fan}}, \ and\ \bibinfo
  {author} {\bibfnamefont {C.-M.}\ \bibnamefont {Hu}},\ }\href {\doibase
  10.1103/PhysRevB.84.054423} {\bibfield  {journal} {\bibinfo  {journal} {Phys.
  Rev. B}\ }\textbf {\bibinfo {volume} {84}},\ \bibinfo {pages} {054423}
  (\bibinfo {year} {2011})}\BibitemShut {NoStop}%
\bibitem [{\citenamefont {Bailey}\ \emph {et~al.}(2013)\citenamefont {Bailey},
  \citenamefont {Cheng}, \citenamefont {Knut}, \citenamefont {Karis},
  \citenamefont {Auffret}, \citenamefont {Zohar}, \citenamefont {Keavney},
  \citenamefont {Warnicke}, \citenamefont {Lee},\ and\ \citenamefont
  {Arena}}]{BaileyNcomm2013}%
  \BibitemOpen
  \bibfield  {author} {\bibinfo {author} {\bibfnamefont {W.~E.}\ \bibnamefont
  {Bailey}}, \bibinfo {author} {\bibfnamefont {C.}~\bibnamefont {Cheng}},
  \bibinfo {author} {\bibfnamefont {R.}~\bibnamefont {Knut}}, \bibinfo {author}
  {\bibfnamefont {O.}~\bibnamefont {Karis}}, \bibinfo {author} {\bibfnamefont
  {S.}~\bibnamefont {Auffret}}, \bibinfo {author} {\bibfnamefont
  {S.}~\bibnamefont {Zohar}}, \bibinfo {author} {\bibfnamefont
  {D.}~\bibnamefont {Keavney}}, \bibinfo {author} {\bibfnamefont
  {P.}~\bibnamefont {Warnicke}}, \bibinfo {author} {\bibfnamefont {J.-S.}\
  \bibnamefont {Lee}}, \ and\ \bibinfo {author} {\bibfnamefont {D.~A.}\
  \bibnamefont {Arena}},\ }\href@noop {} {\bibfield  {journal} {\bibinfo
  {journal} {Nat. Commun.}\ }\textbf {\bibinfo {volume} {4}},\ \bibinfo {pages}
  {2025} (\bibinfo {year} {2013})}\BibitemShut {NoStop}%
\bibitem [{\citenamefont {Vlaminck}\ \emph {et~al.}(2012)\citenamefont
  {Vlaminck}, \citenamefont {Schultheiss}, \citenamefont {Pearson},
  \citenamefont {Fradin}, \citenamefont {Bader},\ and\ \citenamefont
  {Hoffmann}}]{VlaminckAPL2012}%
  \BibitemOpen
  \bibfield  {author} {\bibinfo {author} {\bibfnamefont {V.}~\bibnamefont
  {Vlaminck}}, \bibinfo {author} {\bibfnamefont {H.}~\bibnamefont
  {Schultheiss}}, \bibinfo {author} {\bibfnamefont {J.~E.}\ \bibnamefont
  {Pearson}}, \bibinfo {author} {\bibfnamefont {F.~Y.}\ \bibnamefont {Fradin}},
  \bibinfo {author} {\bibfnamefont {S.~D.}\ \bibnamefont {Bader}}, \ and\
  \bibinfo {author} {\bibfnamefont {A.}~\bibnamefont {Hoffmann}},\ }\href@noop
  {} {\bibfield  {journal} {\bibinfo  {journal} {Appl. Phys. Lett.}\ }\textbf
  {\bibinfo {volume} {101}},\ \bibinfo {pages} {252406} (\bibinfo {year}
  {2012})}\BibitemShut {NoStop}%
\bibitem [{\citenamefont {Qiu}\ and\ \citenamefont
  {Bader}(2000)}]{BaderRSI2000}%
  \BibitemOpen
  \bibfield  {author} {\bibinfo {author} {\bibfnamefont {Z.~Q.}\ \bibnamefont
  {Qiu}}\ and\ \bibinfo {author} {\bibfnamefont {S.~D.}\ \bibnamefont
  {Bader}},\ }\href {\doibase 10.1063/1.1150496} {\bibfield  {journal}
  {\bibinfo  {journal} {Rev. Sci. Instrum.}\ }\textbf {\bibinfo {volume}
  {71}},\ \bibinfo {pages} {1243} (\bibinfo {year} {2000})}\BibitemShut
  {NoStop}%
\bibitem [{\citenamefont {Fan}\ \emph {et~al.}(2016)\citenamefont {Fan},
  \citenamefont {Mellnik}, \citenamefont {Wang}, \citenamefont {Reynolds},
  \citenamefont {Wang}, \citenamefont {Celik}, \citenamefont {Lorenz},
  \citenamefont {Ralph},\ and\ \citenamefont {Xiao}}]{FanAPL2016}%
  \BibitemOpen
  \bibfield  {author} {\bibinfo {author} {\bibfnamefont {X.}~\bibnamefont
  {Fan}}, \bibinfo {author} {\bibfnamefont {A.~R.}\ \bibnamefont {Mellnik}},
  \bibinfo {author} {\bibfnamefont {W.}~\bibnamefont {Wang}}, \bibinfo {author}
  {\bibfnamefont {N.}~\bibnamefont {Reynolds}}, \bibinfo {author}
  {\bibfnamefont {T.}~\bibnamefont {Wang}}, \bibinfo {author} {\bibfnamefont
  {H.}~\bibnamefont {Celik}}, \bibinfo {author} {\bibfnamefont {V.~O.}\
  \bibnamefont {Lorenz}}, \bibinfo {author} {\bibfnamefont {D.~C.}\
  \bibnamefont {Ralph}}, \ and\ \bibinfo {author} {\bibfnamefont {J.~Q.}\
  \bibnamefont {Xiao}},\ }\href@noop {} {\bibfield  {journal} {\bibinfo
  {journal} {Appl. Phys. Lett.}\ }\textbf {\bibinfo {volume} {109}},\ \bibinfo
  {pages} {122406} (\bibinfo {year} {2016})}\BibitemShut {NoStop}%
\bibitem [{\citenamefont {Marui}\ \emph {et~al.}(2018)\citenamefont {Marui},
  \citenamefont {Kawaguchi},\ and\ \citenamefont {Hayashi}}]{HayashiAPEX2018}%
  \BibitemOpen
  \bibfield  {author} {\bibinfo {author} {\bibfnamefont {Y.}~\bibnamefont
  {Marui}}, \bibinfo {author} {\bibfnamefont {M.}~\bibnamefont {Kawaguchi}}, \
  and\ \bibinfo {author} {\bibfnamefont {M.}~\bibnamefont {Hayashi}},\ }\href
  {http://stacks.iop.org/1882-0786/11/i=9/a=093001} {\bibfield  {journal}
  {\bibinfo  {journal} {Appl. Phys. Express}\ }\textbf {\bibinfo {volume}
  {11}},\ \bibinfo {pages} {093001} (\bibinfo {year} {2018})}\BibitemShut
  {NoStop}%
\bibitem [{\citenamefont {Tsai}\ \emph {et~al.}(2018)\citenamefont {Tsai},
  \citenamefont {Chen}, \citenamefont {Wu}, \citenamefont {Chan},\ and\
  \citenamefont {C.-F.}}]{TsaiSREP2018}%
  \BibitemOpen
  \bibfield  {author} {\bibinfo {author} {\bibfnamefont {T.-Y.}\ \bibnamefont
  {Tsai}}, \bibinfo {author} {\bibfnamefont {T.-Y.}\ \bibnamefont {Chen}},
  \bibinfo {author} {\bibfnamefont {C.-T.}\ \bibnamefont {Wu}}, \bibinfo
  {author} {\bibfnamefont {H.-I.}\ \bibnamefont {Chan}}, \ and\ \bibinfo
  {author} {\bibfnamefont {P.}~\bibnamefont {C.-F.}},\ }\href@noop {}
  {\bibfield  {journal} {\bibinfo  {journal} {Sci. Rep.}\ }\textbf {\bibinfo
  {volume} {8}},\ \bibinfo {pages} {5613} (\bibinfo {year} {2018})}\BibitemShut
  {NoStop}%
\bibitem [{\citenamefont {Montazeri}\ \emph {et~al.}(2015)\citenamefont
  {Montazeri}, \citenamefont {Upadhyaya}, \citenamefont {Onbasli},
  \citenamefont {Yu}, \citenamefont {Wong}, \citenamefont {Lang}, \citenamefont
  {Fan}, \citenamefont {Li}, \citenamefont {Khalili-Amiri}, \citenamefont
  {Schwartz}, \citenamefont {Ross},\ and\ \citenamefont
  {Wang}}]{MontazeriNcomm2015}%
  \BibitemOpen
  \bibfield  {author} {\bibinfo {author} {\bibfnamefont {M.}~\bibnamefont
  {Montazeri}}, \bibinfo {author} {\bibfnamefont {P.}~\bibnamefont
  {Upadhyaya}}, \bibinfo {author} {\bibfnamefont {M.~C.}\ \bibnamefont
  {Onbasli}}, \bibinfo {author} {\bibfnamefont {G.}~\bibnamefont {Yu}},
  \bibinfo {author} {\bibfnamefont {K.~L.}\ \bibnamefont {Wong}}, \bibinfo
  {author} {\bibfnamefont {M.}~\bibnamefont {Lang}}, \bibinfo {author}
  {\bibfnamefont {Y.}~\bibnamefont {Fan}}, \bibinfo {author} {\bibfnamefont
  {X.}~\bibnamefont {Li}}, \bibinfo {author} {\bibfnamefont {P.}~\bibnamefont
  {Khalili-Amiri}}, \bibinfo {author} {\bibfnamefont {R.~N.}\ \bibnamefont
  {Schwartz}}, \bibinfo {author} {\bibfnamefont {C.~A.}\ \bibnamefont {Ross}},
  \ and\ \bibinfo {author} {\bibfnamefont {K.~L.}\ \bibnamefont {Wang}},\
  }\href@noop {} {\bibfield  {journal} {\bibinfo  {journal} {Nat. Commun.}\
  }\textbf {\bibinfo {volume} {6}},\ \bibinfo {pages} {8958} (\bibinfo {year}
  {2015})}\BibitemShut {NoStop}%
\bibitem [{\citenamefont {Nembach}\ \emph {et~al.}(2013)\citenamefont
  {Nembach}, \citenamefont {Shaw}, \citenamefont {Boone},\ and\ \citenamefont
  {Silva}}]{nembachPRL2013}%
  \BibitemOpen
  \bibfield  {author} {\bibinfo {author} {\bibfnamefont {H.~T.}\ \bibnamefont
  {Nembach}}, \bibinfo {author} {\bibfnamefont {J.~M.}\ \bibnamefont {Shaw}},
  \bibinfo {author} {\bibfnamefont {C.~T.}\ \bibnamefont {Boone}}, \ and\
  \bibinfo {author} {\bibfnamefont {T.~J.}\ \bibnamefont {Silva}},\ }\href@noop
  {} {\bibfield  {journal} {\bibinfo  {journal} {Phys. Rev. Lett.}\ }\textbf
  {\bibinfo {volume} {110}},\ \bibinfo {pages} {117201} (\bibinfo {year}
  {2013})}\BibitemShut {NoStop}%
\bibitem [{\citenamefont {Moriyama}\ \emph {et~al.}(2015)\citenamefont
  {Moriyama}, \citenamefont {Yoon},\ and\ \citenamefont
  {McMichael}}]{MoriyamaJAP2015}%
  \BibitemOpen
  \bibfield  {author} {\bibinfo {author} {\bibfnamefont {T.}~\bibnamefont
  {Moriyama}}, \bibinfo {author} {\bibfnamefont {S.}~\bibnamefont {Yoon}}, \
  and\ \bibinfo {author} {\bibfnamefont {R.~D.}\ \bibnamefont {McMichael}},\
  }\href {\doibase 10.1063/1.4922126} {\bibfield  {journal} {\bibinfo
  {journal} {J. Appl. Phys.}\ }\textbf {\bibinfo {volume} {117}},\ \bibinfo
  {pages} {213908} (\bibinfo {year} {2015})}\BibitemShut {NoStop}%
\bibitem [{\citenamefont {Guo}\ \emph {et~al.}(2015)\citenamefont {Guo},
  \citenamefont {Bartell}, \citenamefont {Ngai},\ and\ \citenamefont
  {Fuchs}}]{GuoPRAppl2015}%
  \BibitemOpen
  \bibfield  {author} {\bibinfo {author} {\bibfnamefont {F.}~\bibnamefont
  {Guo}}, \bibinfo {author} {\bibfnamefont {J.~M.}\ \bibnamefont {Bartell}},
  \bibinfo {author} {\bibfnamefont {D.~H.}\ \bibnamefont {Ngai}}, \ and\
  \bibinfo {author} {\bibfnamefont {G.~D.}\ \bibnamefont {Fuchs}},\ }\href
  {\doibase 10.1103/PhysRevApplied.4.044004} {\bibfield  {journal} {\bibinfo
  {journal} {Phys. Rev. Applied}\ }\textbf {\bibinfo {volume} {4}},\ \bibinfo
  {pages} {044004} (\bibinfo {year} {2015})}\BibitemShut {NoStop}%
\bibitem [{\citenamefont {Bartell}\ \emph {et~al.}(2015)\citenamefont
  {Bartell}, \citenamefont {Ngai}, \citenamefont {Leng},\ and\ \citenamefont
  {Fuchs}}]{FuchsNcomm2015}%
  \BibitemOpen
  \bibfield  {author} {\bibinfo {author} {\bibfnamefont {J.~M.}\ \bibnamefont
  {Bartell}}, \bibinfo {author} {\bibfnamefont {D.~H.}\ \bibnamefont {Ngai}},
  \bibinfo {author} {\bibfnamefont {Z.}~\bibnamefont {Leng}}, \ and\ \bibinfo
  {author} {\bibfnamefont {G.~D.}\ \bibnamefont {Fuchs}},\ }\href@noop {}
  {\bibfield  {journal} {\bibinfo  {journal} {Nat. Commun.}\ }\textbf {\bibinfo
  {volume} {6}},\ \bibinfo {pages} {8460} (\bibinfo {year} {2015})}\BibitemShut
  {NoStop}%
\bibitem [{\citenamefont {Yoon}\ \emph {et~al.}(2016)\citenamefont {Yoon},
  \citenamefont {Liu},\ and\ \citenamefont {McMichael}}]{YoonPRB2016}%
  \BibitemOpen
  \bibfield  {author} {\bibinfo {author} {\bibfnamefont {S.}~\bibnamefont
  {Yoon}}, \bibinfo {author} {\bibfnamefont {J.}~\bibnamefont {Liu}}, \ and\
  \bibinfo {author} {\bibfnamefont {R.~D.}\ \bibnamefont {McMichael}},\ }\href
  {\doibase 10.1103/PhysRevB.93.144423} {\bibfield  {journal} {\bibinfo
  {journal} {Phys. Rev. B}\ }\textbf {\bibinfo {volume} {93}},\ \bibinfo
  {pages} {144423} (\bibinfo {year} {2016})}\BibitemShut {NoStop}%
\bibitem [{\citenamefont {Mosendz}\ \emph
  {et~al.}(2010{\natexlab{b}})\citenamefont {Mosendz}, \citenamefont {Pearson},
  \citenamefont {Fradin}, \citenamefont {Bader},\ and\ \citenamefont
  {Hoffmann}}]{MosendzAPL2010}%
  \BibitemOpen
  \bibfield  {author} {\bibinfo {author} {\bibfnamefont {O.}~\bibnamefont
  {Mosendz}}, \bibinfo {author} {\bibfnamefont {J.~E.}\ \bibnamefont
  {Pearson}}, \bibinfo {author} {\bibfnamefont {F.~Y.}\ \bibnamefont {Fradin}},
  \bibinfo {author} {\bibfnamefont {S.~D.}\ \bibnamefont {Bader}}, \ and\
  \bibinfo {author} {\bibfnamefont {A.}~\bibnamefont {Hoffmann}},\ }\href@noop
  {} {\bibfield  {journal} {\bibinfo  {journal} {Appl. Phys. Lett.}\ }\textbf
  {\bibinfo {volume} {96}},\ \bibinfo {pages} {022502} (\bibinfo {year}
  {2010}{\natexlab{b}})}\BibitemShut {NoStop}%
\bibitem [{\citenamefont {Jungfleisch}\ \emph {et~al.}(2016)\citenamefont
  {Jungfleisch}, \citenamefont {Zhang}, \citenamefont {Sklenar}, \citenamefont
  {Ding}, \citenamefont {Jiang}, \citenamefont {Chang}, \citenamefont {Fradin},
  \citenamefont {Pearson}, \citenamefont {Ketterson}, \citenamefont {Novosad},
  \citenamefont {Wu},\ and\ \citenamefont {Hoffmann}}]{JungfleischPRL2016}%
  \BibitemOpen
  \bibfield  {author} {\bibinfo {author} {\bibfnamefont {M.~B.}\ \bibnamefont
  {Jungfleisch}}, \bibinfo {author} {\bibfnamefont {W.}~\bibnamefont {Zhang}},
  \bibinfo {author} {\bibfnamefont {J.}~\bibnamefont {Sklenar}}, \bibinfo
  {author} {\bibfnamefont {J.}~\bibnamefont {Ding}}, \bibinfo {author}
  {\bibfnamefont {W.}~\bibnamefont {Jiang}}, \bibinfo {author} {\bibfnamefont
  {H.}~\bibnamefont {Chang}}, \bibinfo {author} {\bibfnamefont {F.~Y.}\
  \bibnamefont {Fradin}}, \bibinfo {author} {\bibfnamefont {J.~E.}\
  \bibnamefont {Pearson}}, \bibinfo {author} {\bibfnamefont {J.~B.}\
  \bibnamefont {Ketterson}}, \bibinfo {author} {\bibfnamefont {V.}~\bibnamefont
  {Novosad}}, \bibinfo {author} {\bibfnamefont {M.}~\bibnamefont {Wu}}, \ and\
  \bibinfo {author} {\bibfnamefont {A.}~\bibnamefont {Hoffmann}},\ }\href
  {\doibase 10.1103/PhysRevLett.116.057601} {\bibfield  {journal} {\bibinfo
  {journal} {Phys. Rev. Lett.}\ }\textbf {\bibinfo {volume} {116}},\ \bibinfo
  {pages} {057601} (\bibinfo {year} {2016})}\BibitemShut {NoStop}%
\bibitem [{\citenamefont {Li}\ \emph {et~al.}(2016{\natexlab{a}})\citenamefont
  {Li}, \citenamefont {Zhang}, \citenamefont {Ding}, \citenamefont {Pearson},
  \citenamefont {Novosad},\ and\ \citenamefont {Hoffmann}}]{LiNanoscale2016}%
  \BibitemOpen
  \bibfield  {author} {\bibinfo {author} {\bibfnamefont {S.}~\bibnamefont
  {Li}}, \bibinfo {author} {\bibfnamefont {W.}~\bibnamefont {Zhang}}, \bibinfo
  {author} {\bibfnamefont {J.}~\bibnamefont {Ding}}, \bibinfo {author}
  {\bibfnamefont {J.~E.}\ \bibnamefont {Pearson}}, \bibinfo {author}
  {\bibfnamefont {V.}~\bibnamefont {Novosad}}, \ and\ \bibinfo {author}
  {\bibfnamefont {A.}~\bibnamefont {Hoffmann}},\ }\href@noop {} {\bibfield
  {journal} {\bibinfo  {journal} {Nanoscale}\ }\textbf {\bibinfo {volume}
  {8}},\ \bibinfo {pages} {388} (\bibinfo {year}
  {2016}{\natexlab{a}})}\BibitemShut {NoStop}%
\bibitem [{\citenamefont {Du}\ \emph {et~al.}(2014)\citenamefont {Du},
  \citenamefont {Wang}, \citenamefont {Yang},\ and\ \citenamefont
  {Hammel}}]{DuCHPRApplied2014}%
  \BibitemOpen
  \bibfield  {author} {\bibinfo {author} {\bibfnamefont {C.}~\bibnamefont
  {Du}}, \bibinfo {author} {\bibfnamefont {H.}~\bibnamefont {Wang}}, \bibinfo
  {author} {\bibfnamefont {F.}~\bibnamefont {Yang}}, \ and\ \bibinfo {author}
  {\bibfnamefont {P.~C.}\ \bibnamefont {Hammel}},\ }\href@noop {} {\bibfield
  {journal} {\bibinfo  {journal} {Phys. Rev. Applied}\ }\textbf {\bibinfo
  {volume} {1}},\ \bibinfo {pages} {044004} (\bibinfo {year}
  {2014})}\BibitemShut {NoStop}%
\bibitem [{\citenamefont {Choi}\ and\ \citenamefont
  {Cahill}(2014)}]{ChoiPRB2015}%
  \BibitemOpen
  \bibfield  {author} {\bibinfo {author} {\bibfnamefont {G.-M.}\ \bibnamefont
  {Choi}}\ and\ \bibinfo {author} {\bibfnamefont {D.~G.}\ \bibnamefont
  {Cahill}},\ }\href {\doibase 10.1103/PhysRevB.90.214432} {\bibfield
  {journal} {\bibinfo  {journal} {Phys. Rev. B}\ }\textbf {\bibinfo {volume}
  {90}},\ \bibinfo {pages} {214432} (\bibinfo {year} {2014})}\BibitemShut
  {NoStop}%
\bibitem [{\citenamefont {Choi}\ \emph {et~al.}(2015)\citenamefont {Choi},
  \citenamefont {Moon}, \citenamefont {Min}, \citenamefont {Lee},\ and\
  \citenamefont {Cahill}}]{ChoiNaturePhys2015}%
  \BibitemOpen
  \bibfield  {author} {\bibinfo {author} {\bibfnamefont {G.-M.}\ \bibnamefont
  {Choi}}, \bibinfo {author} {\bibfnamefont {C.-H.}\ \bibnamefont {Moon}},
  \bibinfo {author} {\bibfnamefont {B.-C.}\ \bibnamefont {Min}}, \bibinfo
  {author} {\bibfnamefont {K.-J.}\ \bibnamefont {Lee}}, \ and\ \bibinfo
  {author} {\bibfnamefont {D.~G.}\ \bibnamefont {Cahill}},\ }\href@noop {}
  {\bibfield  {journal} {\bibinfo  {journal} {Nature Phys.}\ }\textbf {\bibinfo
  {volume} {11}},\ \bibinfo {pages} {576} (\bibinfo {year} {2015})}\BibitemShut
  {NoStop}%
\bibitem [{\citenamefont {Liu}\ \emph {et~al.}(2012)\citenamefont {Liu},
  \citenamefont {Pai}, \citenamefont {Li}, \citenamefont {Tseng}, \citenamefont
  {Ralph},\ and\ \citenamefont {Buhrman}}]{LiuScience2012}%
  \BibitemOpen
  \bibfield  {author} {\bibinfo {author} {\bibfnamefont {L.}~\bibnamefont
  {Liu}}, \bibinfo {author} {\bibfnamefont {C.-F.}\ \bibnamefont {Pai}},
  \bibinfo {author} {\bibfnamefont {Y.}~\bibnamefont {Li}}, \bibinfo {author}
  {\bibfnamefont {H.~W.}\ \bibnamefont {Tseng}}, \bibinfo {author}
  {\bibfnamefont {D.~C.}\ \bibnamefont {Ralph}}, \ and\ \bibinfo {author}
  {\bibfnamefont {R.~A.}\ \bibnamefont {Buhrman}},\ }\href {\doibase
  10.1126/science.1218197} {\bibfield  {journal} {\bibinfo  {journal}
  {Science}\ }\textbf {\bibinfo {volume} {336}},\ \bibinfo {pages} {555}
  (\bibinfo {year} {2012})}\BibitemShut {NoStop}%
\bibitem [{\citenamefont {Hahn}\ \emph {et~al.}(2013)\citenamefont {Hahn},
  \citenamefont {de~Loubens}, \citenamefont {Klein}, \citenamefont {Viret},
  \citenamefont {Naletov},\ and\ \citenamefont {Ben~Youssef}}]{HahnPRB2013}%
  \BibitemOpen
  \bibfield  {author} {\bibinfo {author} {\bibfnamefont {C.}~\bibnamefont
  {Hahn}}, \bibinfo {author} {\bibfnamefont {G.}~\bibnamefont {de~Loubens}},
  \bibinfo {author} {\bibfnamefont {O.}~\bibnamefont {Klein}}, \bibinfo
  {author} {\bibfnamefont {M.}~\bibnamefont {Viret}}, \bibinfo {author}
  {\bibfnamefont {V.~V.}\ \bibnamefont {Naletov}}, \ and\ \bibinfo {author}
  {\bibfnamefont {J.}~\bibnamefont {Ben~Youssef}},\ }\href {\doibase
  10.1103/PhysRevB.87.174417} {\bibfield  {journal} {\bibinfo  {journal} {Phys.
  Rev. B}\ }\textbf {\bibinfo {volume} {87}},\ \bibinfo {pages} {174417}
  (\bibinfo {year} {2013})}\BibitemShut {NoStop}%
\bibitem [{\citenamefont {Wang}\ \emph {et~al.}(2014)\citenamefont {Wang},
  \citenamefont {Du}, \citenamefont {Pu}, \citenamefont {Adur}, \citenamefont
  {Hammel},\ and\ \citenamefont {Yang}}]{WangPRL2014}%
  \BibitemOpen
  \bibfield  {author} {\bibinfo {author} {\bibfnamefont {H.~L.}\ \bibnamefont
  {Wang}}, \bibinfo {author} {\bibfnamefont {C.~H.}\ \bibnamefont {Du}},
  \bibinfo {author} {\bibfnamefont {Y.}~\bibnamefont {Pu}}, \bibinfo {author}
  {\bibfnamefont {R.}~\bibnamefont {Adur}}, \bibinfo {author} {\bibfnamefont
  {P.~C.}\ \bibnamefont {Hammel}}, \ and\ \bibinfo {author} {\bibfnamefont
  {F.~Y.}\ \bibnamefont {Yang}},\ }\href {\doibase
  10.1103/PhysRevLett.112.197201} {\bibfield  {journal} {\bibinfo  {journal}
  {Phys. Rev. Lett.}\ }\textbf {\bibinfo {volume} {112}},\ \bibinfo {pages}
  {197201} (\bibinfo {year} {2014})}\BibitemShut {NoStop}%
\bibitem [{\citenamefont {Slonczewski}(1996)}]{slonczewskiJMMM1996}%
  \BibitemOpen
  \bibfield  {author} {\bibinfo {author} {\bibfnamefont {J.~C.}\ \bibnamefont
  {Slonczewski}},\ }\href@noop {} {\bibfield  {journal} {\bibinfo  {journal}
  {J. Magn. Magn. Mater}\ }\textbf {\bibinfo {volume} {159}},\ \bibinfo {pages}
  {L1} (\bibinfo {year} {1996})}\BibitemShut {NoStop}%
\bibitem [{\citenamefont {Zhang}\ \emph {et~al.}(2015)\citenamefont {Zhang},
  \citenamefont {Jungfleisch}, \citenamefont {Freimuth}, \citenamefont {Jiang},
  \citenamefont {Sklenar}, \citenamefont {Pearson}, \citenamefont {Ketterson},
  \citenamefont {Mokrousov},\ and\ \citenamefont {Hoffmann}}]{WeiZhangPRB2015}%
  \BibitemOpen
  \bibfield  {author} {\bibinfo {author} {\bibfnamefont {W.}~\bibnamefont
  {Zhang}}, \bibinfo {author} {\bibfnamefont {M.~B.}\ \bibnamefont
  {Jungfleisch}}, \bibinfo {author} {\bibfnamefont {F.}~\bibnamefont
  {Freimuth}}, \bibinfo {author} {\bibfnamefont {W.}~\bibnamefont {Jiang}},
  \bibinfo {author} {\bibfnamefont {J.}~\bibnamefont {Sklenar}}, \bibinfo
  {author} {\bibfnamefont {J.~E.}\ \bibnamefont {Pearson}}, \bibinfo {author}
  {\bibfnamefont {J.~B.}\ \bibnamefont {Ketterson}}, \bibinfo {author}
  {\bibfnamefont {Y.}~\bibnamefont {Mokrousov}}, \ and\ \bibinfo {author}
  {\bibfnamefont {A.}~\bibnamefont {Hoffmann}},\ }\href {\doibase
  10.1103/PhysRevB.92.144405} {\bibfield  {journal} {\bibinfo  {journal} {Phys.
  Rev. B}\ }\textbf {\bibinfo {volume} {92}},\ \bibinfo {pages} {144405}
  (\bibinfo {year} {2015})}\BibitemShut {NoStop}%
\bibitem [{\citenamefont {Guo}\ \emph {et~al.}(2013)\citenamefont {Guo},
  \citenamefont {Belova},\ and\ \citenamefont {McMichael}}]{GuoPRL2013}%
  \BibitemOpen
  \bibfield  {author} {\bibinfo {author} {\bibfnamefont {F.}~\bibnamefont
  {Guo}}, \bibinfo {author} {\bibfnamefont {L.~M.}\ \bibnamefont {Belova}}, \
  and\ \bibinfo {author} {\bibfnamefont {R.~D.}\ \bibnamefont {McMichael}},\
  }\href@noop {} {\bibfield  {journal} {\bibinfo  {journal} {Phys. Rev. Lett.}\
  }\textbf {\bibinfo {volume} {110}},\ \bibinfo {pages} {017601} (\bibinfo
  {year} {2013})}\BibitemShut {NoStop}%
\bibitem [{\citenamefont {Jiang}\ \emph {et~al.}(2018)\citenamefont {Jiang},
  \citenamefont {Chung}, \citenamefont {Le}, \citenamefont {Mazraati},
  \citenamefont {Houshang},\ and\ \citenamefont
  {\r{A}kerman}}]{JiangPRApplied2018}%
  \BibitemOpen
  \bibfield  {author} {\bibinfo {author} {\bibfnamefont {S.}~\bibnamefont
  {Jiang}}, \bibinfo {author} {\bibfnamefont {S.}~\bibnamefont {Chung}},
  \bibinfo {author} {\bibfnamefont {Q.~T.}\ \bibnamefont {Le}}, \bibinfo
  {author} {\bibfnamefont {H.}~\bibnamefont {Mazraati}}, \bibinfo {author}
  {\bibfnamefont {A.}~\bibnamefont {Houshang}}, \ and\ \bibinfo {author}
  {\bibfnamefont {J.}~\bibnamefont {\r{A}kerman}},\ }\href@noop {} {\bibfield
  {journal} {\bibinfo  {journal} {Phys. Rev. Applied}\ }\textbf {\bibinfo
  {volume} {10}},\ \bibinfo {pages} {054014} (\bibinfo {year}
  {2018})}\BibitemShut {NoStop}%
\bibitem [{\citenamefont {Zhao}\ \emph {et~al.}(2018)\citenamefont {Zhao},
  \citenamefont {Wang}, \citenamefont {Zhou}, \citenamefont {Li}, \citenamefont
  {Dong}, \citenamefont {Zhang}, \citenamefont {Peng}, \citenamefont {Min},
  \citenamefont {Hu}, \citenamefont {Ma}, \citenamefont {Ren}, \citenamefont
  {Ye}, \citenamefont {Chen}, \citenamefont {Yu}, \citenamefont {Nan},\ and\
  \citenamefont {Liu}}]{ZhouZY_AM_2018}%
  \BibitemOpen
  \bibfield  {author} {\bibinfo {author} {\bibfnamefont {S.}~\bibnamefont
  {Zhao}}, \bibinfo {author} {\bibfnamefont {L.}~\bibnamefont {Wang}}, \bibinfo
  {author} {\bibfnamefont {Z.}~\bibnamefont {Zhou}}, \bibinfo {author}
  {\bibfnamefont {C.}~\bibnamefont {Li}}, \bibinfo {author} {\bibfnamefont
  {G.}~\bibnamefont {Dong}}, \bibinfo {author} {\bibfnamefont {L.}~\bibnamefont
  {Zhang}}, \bibinfo {author} {\bibfnamefont {B.}~\bibnamefont {Peng}},
  \bibinfo {author} {\bibfnamefont {T.}~\bibnamefont {Min}}, \bibinfo {author}
  {\bibfnamefont {Z.}~\bibnamefont {Hu}}, \bibinfo {author} {\bibfnamefont
  {J.}~\bibnamefont {Ma}}, \bibinfo {author} {\bibfnamefont {W.}~\bibnamefont
  {Ren}}, \bibinfo {author} {\bibfnamefont {Z.}~\bibnamefont {Ye}}, \bibinfo
  {author} {\bibfnamefont {W.}~\bibnamefont {Chen}}, \bibinfo {author}
  {\bibfnamefont {P.}~\bibnamefont {Yu}}, \bibinfo {author} {\bibfnamefont
  {C.}~\bibnamefont {Nan}}, \ and\ \bibinfo {author} {\bibfnamefont
  {M.}~\bibnamefont {Liu}},\ }\href@noop {} {\bibfield  {journal} {\bibinfo
  {journal} {Adv. Mater.}\ }\textbf {\bibinfo {volume} {30}},\ \bibinfo {pages}
  {1801639} (\bibinfo {year} {2018})}\BibitemShut {NoStop}%
\bibitem [{\citenamefont {Yang}\ \emph {et~al.}(2018)\citenamefont {Yang},
  \citenamefont {Wang}, \citenamefont {Zhou}, \citenamefont {Wang},
  \citenamefont {Zhang}, \citenamefont {Zhao}, \citenamefont {Dong},
  \citenamefont {Cheng}, \citenamefont {Min}, \citenamefont {Hu}, \citenamefont
  {Chen}, \citenamefont {Xia},\ and\ \citenamefont {Liu}}]{ZhouZY_NCOMM_2018}%
  \BibitemOpen
  \bibfield  {author} {\bibinfo {author} {\bibfnamefont {Q.}~\bibnamefont
  {Yang}}, \bibinfo {author} {\bibfnamefont {L.}~\bibnamefont {Wang}}, \bibinfo
  {author} {\bibfnamefont {Z.}~\bibnamefont {Zhou}}, \bibinfo {author}
  {\bibfnamefont {L.}~\bibnamefont {Wang}}, \bibinfo {author} {\bibfnamefont
  {Y.}~\bibnamefont {Zhang}}, \bibinfo {author} {\bibfnamefont
  {S.}~\bibnamefont {Zhao}}, \bibinfo {author} {\bibfnamefont {G.}~\bibnamefont
  {Dong}}, \bibinfo {author} {\bibfnamefont {Y.}~\bibnamefont {Cheng}},
  \bibinfo {author} {\bibfnamefont {T.}~\bibnamefont {Min}}, \bibinfo {author}
  {\bibfnamefont {Z.}~\bibnamefont {Hu}}, \bibinfo {author} {\bibfnamefont
  {W.}~\bibnamefont {Chen}}, \bibinfo {author} {\bibfnamefont {K.}~\bibnamefont
  {Xia}}, \ and\ \bibinfo {author} {\bibfnamefont {M.}~\bibnamefont {Liu}},\
  }\href@noop {} {\bibfield  {journal} {\bibinfo  {journal} {Nature Commun}\
  }\textbf {\bibinfo {volume} {9}},\ \bibinfo {pages} {991} (\bibinfo {year}
  {2018})}\BibitemShut {NoStop}%
\bibitem [{\citenamefont {Jungfleisch}\ \emph {et~al.}(2017)\citenamefont
  {Jungfleisch}, \citenamefont {Ding}, \citenamefont {Zhang}, \citenamefont
  {Jiang}, \citenamefont {Pearson}, \citenamefont {Novosad},\ and\
  \citenamefont {Hoffmann}}]{JungfleischNanoLett2017}%
  \BibitemOpen
  \bibfield  {author} {\bibinfo {author} {\bibfnamefont {M.~B.}\ \bibnamefont
  {Jungfleisch}}, \bibinfo {author} {\bibfnamefont {J.}~\bibnamefont {Ding}},
  \bibinfo {author} {\bibfnamefont {W.}~\bibnamefont {Zhang}}, \bibinfo
  {author} {\bibfnamefont {W.}~\bibnamefont {Jiang}}, \bibinfo {author}
  {\bibfnamefont {J.~E.}\ \bibnamefont {Pearson}}, \bibinfo {author}
  {\bibfnamefont {V.}~\bibnamefont {Novosad}}, \ and\ \bibinfo {author}
  {\bibfnamefont {A.}~\bibnamefont {Hoffmann}},\ }\href {\doibase
  10.1021/acs.nanolett.6b02794} {\bibfield  {journal} {\bibinfo  {journal}
  {Nano Letters}\ }\textbf {\bibinfo {volume} {17}},\ \bibinfo {pages} {8}
  (\bibinfo {year} {2017})}\BibitemShut {NoStop}%
\bibitem [{\citenamefont {Sklenar}\ \emph {et~al.}(2015)\citenamefont
  {Sklenar}, \citenamefont {Zhang}, \citenamefont {Jungfleisch}, \citenamefont
  {Jiang}, \citenamefont {Chang}, \citenamefont {Pearson}, \citenamefont {Wu},
  \citenamefont {Ketterson},\ and\ \citenamefont {Hoffmann}}]{SklenarPRB2015}%
  \BibitemOpen
  \bibfield  {author} {\bibinfo {author} {\bibfnamefont {J.}~\bibnamefont
  {Sklenar}}, \bibinfo {author} {\bibfnamefont {W.}~\bibnamefont {Zhang}},
  \bibinfo {author} {\bibfnamefont {M.~B.}\ \bibnamefont {Jungfleisch}},
  \bibinfo {author} {\bibfnamefont {W.}~\bibnamefont {Jiang}}, \bibinfo
  {author} {\bibfnamefont {H.}~\bibnamefont {Chang}}, \bibinfo {author}
  {\bibfnamefont {J.~E.}\ \bibnamefont {Pearson}}, \bibinfo {author}
  {\bibfnamefont {M.}~\bibnamefont {Wu}}, \bibinfo {author} {\bibfnamefont
  {J.~B.}\ \bibnamefont {Ketterson}}, \ and\ \bibinfo {author} {\bibfnamefont
  {A.}~\bibnamefont {Hoffmann}},\ }\href@noop {} {\bibfield  {journal}
  {\bibinfo  {journal} {Phys. Rev. B}\ }\textbf {\bibinfo {volume} {92}},\
  \bibinfo {pages} {174406} (\bibinfo {year} {2015})}\BibitemShut {NoStop}%
\bibitem [{\citenamefont {Schreier}\ \emph {et~al.}(2015)\citenamefont
  {Schreier}, \citenamefont {Chiba}, \citenamefont {Niedermayr}, \citenamefont
  {Lotze}, \citenamefont {Huebl}, \citenamefont {Gepr\"{a}gs}, \citenamefont
  {Takahashi}, \citenamefont {Bauer}, \citenamefont {Gross},\ and\
  \citenamefont {Goennenwein}}]{SchreierPRB2015}%
  \BibitemOpen
  \bibfield  {author} {\bibinfo {author} {\bibfnamefont {M.}~\bibnamefont
  {Schreier}}, \bibinfo {author} {\bibfnamefont {T.}~\bibnamefont {Chiba}},
  \bibinfo {author} {\bibfnamefont {A.}~\bibnamefont {Niedermayr}}, \bibinfo
  {author} {\bibfnamefont {J.}~\bibnamefont {Lotze}}, \bibinfo {author}
  {\bibfnamefont {H.}~\bibnamefont {Huebl}}, \bibinfo {author} {\bibfnamefont
  {S.}~\bibnamefont {Gepr\"{a}gs}}, \bibinfo {author} {\bibfnamefont
  {S.}~\bibnamefont {Takahashi}}, \bibinfo {author} {\bibfnamefont {G.~E.~W.}\
  \bibnamefont {Bauer}}, \bibinfo {author} {\bibfnamefont {R.}~\bibnamefont
  {Gross}}, \ and\ \bibinfo {author} {\bibfnamefont {S.~T.~B.}\ \bibnamefont
  {Goennenwein}},\ }\href@noop {} {\bibfield  {journal} {\bibinfo  {journal}
  {Phys. Rev. B}\ }\textbf {\bibinfo {volume} {92}},\ \bibinfo {pages} {144411}
  (\bibinfo {year} {2015})}\BibitemShut {NoStop}%
\bibitem [{\citenamefont {Saidaoui}\ and\ \citenamefont
  {Manchon}(2016)}]{MohamedPRL2016}%
  \BibitemOpen
  \bibfield  {author} {\bibinfo {author} {\bibfnamefont {H.~B.~M.}\
  \bibnamefont {Saidaoui}}\ and\ \bibinfo {author} {\bibfnamefont
  {A.}~\bibnamefont {Manchon}},\ }\href@noop {} {\bibfield  {journal} {\bibinfo
   {journal} {Phys. Rev. Lett.}\ }\textbf {\bibinfo {volume} {117}},\ \bibinfo
  {pages} {036601} (\bibinfo {year} {2016})}\BibitemShut {NoStop}%
\bibitem [{\citenamefont {Gibbons}\ \emph {et~al.}(2018)\citenamefont
  {Gibbons}, \citenamefont {MacNeill}, \citenamefont {Buhrman},\ and\
  \citenamefont {Ralph}}]{GibbonsPRApplied2018}%
  \BibitemOpen
  \bibfield  {author} {\bibinfo {author} {\bibfnamefont {J.~D.}\ \bibnamefont
  {Gibbons}}, \bibinfo {author} {\bibfnamefont {D.}~\bibnamefont {MacNeill}},
  \bibinfo {author} {\bibfnamefont {R.~A.}\ \bibnamefont {Buhrman}}, \ and\
  \bibinfo {author} {\bibfnamefont {D.~C.}\ \bibnamefont {Ralph}},\ }\href@noop
  {} {\bibfield  {journal} {\bibinfo  {journal} {Phys. Rev. Applied}\ }\textbf
  {\bibinfo {volume} {9}},\ \bibinfo {pages} {064033} (\bibinfo {year}
  {2018})}\BibitemShut {NoStop}%
\bibitem [{\citenamefont {Amin}\ \emph {et~al.}(2018)\citenamefont {Amin},
  \citenamefont {Zemen},\ and\ \citenamefont {Stiles}}]{AminPRL2018}%
  \BibitemOpen
  \bibfield  {author} {\bibinfo {author} {\bibfnamefont {V.~P.}\ \bibnamefont
  {Amin}}, \bibinfo {author} {\bibfnamefont {J.}~\bibnamefont {Zemen}}, \ and\
  \bibinfo {author} {\bibfnamefont {M.~D.}\ \bibnamefont {Stiles}},\
  }\href@noop {} {\bibfield  {journal} {\bibinfo  {journal} {Phys. Rev. Lett.}\
  }\textbf {\bibinfo {volume} {121}},\ \bibinfo {pages} {136805} (\bibinfo
  {year} {2018})}\BibitemShut {NoStop}%
\bibitem [{\citenamefont {Baek}\ \emph {et~al.}(2018)\citenamefont {Baek},
  \citenamefont {Amin}, \citenamefont {Oh}, \citenamefont {Go}, \citenamefont
  {Lee}, \citenamefont {Lee}, \citenamefont {Kim}, \citenamefont {Stiles},
  \citenamefont {Park},\ and\ \citenamefont {Lee}}]{BaekNatureMater2018}%
  \BibitemOpen
  \bibfield  {author} {\bibinfo {author} {\bibfnamefont {S.-H.~C.}\
  \bibnamefont {Baek}}, \bibinfo {author} {\bibfnamefont {V.~P.}\ \bibnamefont
  {Amin}}, \bibinfo {author} {\bibfnamefont {Y.-W.}\ \bibnamefont {Oh}},
  \bibinfo {author} {\bibfnamefont {G.}~\bibnamefont {Go}}, \bibinfo {author}
  {\bibfnamefont {S.-J.}\ \bibnamefont {Lee}}, \bibinfo {author} {\bibfnamefont
  {G.-H.}\ \bibnamefont {Lee}}, \bibinfo {author} {\bibfnamefont {K.-J.}\
  \bibnamefont {Kim}}, \bibinfo {author} {\bibfnamefont {M.~D.}\ \bibnamefont
  {Stiles}}, \bibinfo {author} {\bibfnamefont {B.-G.}\ \bibnamefont {Park}}, \
  and\ \bibinfo {author} {\bibfnamefont {K.-J.}\ \bibnamefont {Lee}},\
  }\href@noop {} {\bibfield  {journal} {\bibinfo  {journal} {Nature Mater}\
  }\textbf {\bibinfo {volume} {17}},\ \bibinfo {pages} {509} (\bibinfo {year}
  {2018})}\BibitemShut {NoStop}%
\bibitem [{\citenamefont {Serga}\ \emph {et~al.}(2006)\citenamefont {Serga},
  \citenamefont {Schneider}, \citenamefont {Hillebrands}, \citenamefont
  {Demokritov},\ and\ \citenamefont {Kostylev}}]{SergaAPL2006}%
  \BibitemOpen
  \bibfield  {author} {\bibinfo {author} {\bibfnamefont {A.~A.}\ \bibnamefont
  {Serga}}, \bibinfo {author} {\bibfnamefont {T.}~\bibnamefont {Schneider}},
  \bibinfo {author} {\bibfnamefont {B.}~\bibnamefont {Hillebrands}}, \bibinfo
  {author} {\bibfnamefont {S.~O.}\ \bibnamefont {Demokritov}}, \ and\ \bibinfo
  {author} {\bibfnamefont {M.~P.}\ \bibnamefont {Kostylev}},\ }\href {\doibase
  10.1063/1.2335627} {\bibfield  {journal} {\bibinfo  {journal} {Appl. Phys.
  Lett.}\ }\textbf {\bibinfo {volume} {89}},\ \bibinfo {pages} {063506}
  (\bibinfo {year} {2006})}\BibitemShut {NoStop}%
\bibitem [{\citenamefont {Arena}\ \emph {et~al.}(2006)\citenamefont {Arena},
  \citenamefont {Vescovo}, \citenamefont {Kao}, \citenamefont {Guan},\ and\
  \citenamefont {Bailey}}]{ArenaPRB2006}%
  \BibitemOpen
  \bibfield  {author} {\bibinfo {author} {\bibfnamefont {D.~A.}\ \bibnamefont
  {Arena}}, \bibinfo {author} {\bibfnamefont {E.}~\bibnamefont {Vescovo}},
  \bibinfo {author} {\bibfnamefont {C.-C.}\ \bibnamefont {Kao}}, \bibinfo
  {author} {\bibfnamefont {Y.}~\bibnamefont {Guan}}, \ and\ \bibinfo {author}
  {\bibfnamefont {W.~E.}\ \bibnamefont {Bailey}},\ }\href {\doibase
  10.1103/PhysRevB.74.064409} {\bibfield  {journal} {\bibinfo  {journal} {Phys.
  Rev. B}\ }\textbf {\bibinfo {volume} {74}},\ \bibinfo {pages} {064409}
  (\bibinfo {year} {2006})}\BibitemShut {NoStop}%
\bibitem [{\citenamefont {Li}\ \emph {et~al.}(2016{\natexlab{b}})\citenamefont
  {Li}, \citenamefont {Shelford}, \citenamefont {Shafer}, \citenamefont {Tan},
  \citenamefont {Deng}, \citenamefont {Keatley}, \citenamefont {Hwang},
  \citenamefont {Arenholz}, \citenamefont {van~der Laan}, \citenamefont
  {Hicken},\ and\ \citenamefont {Qiu}}]{ZQQiuPRL2016}%
  \BibitemOpen
  \bibfield  {author} {\bibinfo {author} {\bibfnamefont {J.}~\bibnamefont
  {Li}}, \bibinfo {author} {\bibfnamefont {L.~R.}\ \bibnamefont {Shelford}},
  \bibinfo {author} {\bibfnamefont {P.}~\bibnamefont {Shafer}}, \bibinfo
  {author} {\bibfnamefont {A.}~\bibnamefont {Tan}}, \bibinfo {author}
  {\bibfnamefont {J.~X.}\ \bibnamefont {Deng}}, \bibinfo {author}
  {\bibfnamefont {P.~S.}\ \bibnamefont {Keatley}}, \bibinfo {author}
  {\bibfnamefont {C.}~\bibnamefont {Hwang}}, \bibinfo {author} {\bibfnamefont
  {E.}~\bibnamefont {Arenholz}}, \bibinfo {author} {\bibfnamefont
  {G.}~\bibnamefont {van~der Laan}}, \bibinfo {author} {\bibfnamefont {R.~J.}\
  \bibnamefont {Hicken}}, \ and\ \bibinfo {author} {\bibfnamefont {Z.~Q.}\
  \bibnamefont {Qiu}},\ }\href {\doibase 10.1103/PhysRevLett.117.076602}
  {\bibfield  {journal} {\bibinfo  {journal} {Phys. Rev. Lett.}\ }\textbf
  {\bibinfo {volume} {117}},\ \bibinfo {pages} {076602} (\bibinfo {year}
  {2016}{\natexlab{b}})}\BibitemShut {NoStop}%
\bibitem [{\citenamefont {Fan}\ \emph {et~al.}(2014)\citenamefont {Fan},
  \citenamefont {Celik}, \citenamefont {Wu}, \citenamefont {Ni}, \citenamefont
  {Lee}, \citenamefont {Lorenz},\ and\ \citenamefont {Xiao}}]{FanNcomm2014}%
  \BibitemOpen
  \bibfield  {author} {\bibinfo {author} {\bibfnamefont {X.}~\bibnamefont
  {Fan}}, \bibinfo {author} {\bibfnamefont {H.}~\bibnamefont {Celik}}, \bibinfo
  {author} {\bibfnamefont {J.}~\bibnamefont {Wu}}, \bibinfo {author}
  {\bibfnamefont {C.}~\bibnamefont {Ni}}, \bibinfo {author} {\bibfnamefont
  {K.-J.}\ \bibnamefont {Lee}}, \bibinfo {author} {\bibfnamefont {V.~O.}\
  \bibnamefont {Lorenz}}, \ and\ \bibinfo {author} {\bibfnamefont {J.~Q.}\
  \bibnamefont {Xiao}},\ }\href@noop {} {\bibfield  {journal} {\bibinfo
  {journal} {Nat. Commun.}\ }\textbf {\bibinfo {volume} {5}},\ \bibinfo {pages}
  {3042} (\bibinfo {year} {2014})}\BibitemShut {NoStop}%
\bibitem [{\citenamefont {Nan}\ \emph {et~al.}(2015)\citenamefont {Nan},
  \citenamefont {Emori}, \citenamefont {Boone}, \citenamefont {Wang},
  \citenamefont {Oxholm}, \citenamefont {Jones}, \citenamefont {Howe},
  \citenamefont {Brown},\ and\ \citenamefont {Sun}}]{NanPRB2015}%
  \BibitemOpen
  \bibfield  {author} {\bibinfo {author} {\bibfnamefont {T.}~\bibnamefont
  {Nan}}, \bibinfo {author} {\bibfnamefont {S.}~\bibnamefont {Emori}}, \bibinfo
  {author} {\bibfnamefont {C.~T.}\ \bibnamefont {Boone}}, \bibinfo {author}
  {\bibfnamefont {X.}~\bibnamefont {Wang}}, \bibinfo {author} {\bibfnamefont
  {T.~M.}\ \bibnamefont {Oxholm}}, \bibinfo {author} {\bibfnamefont {J.~G.}\
  \bibnamefont {Jones}}, \bibinfo {author} {\bibfnamefont {B.~M.}\ \bibnamefont
  {Howe}}, \bibinfo {author} {\bibfnamefont {G.~J.}\ \bibnamefont {Brown}}, \
  and\ \bibinfo {author} {\bibfnamefont {N.~X.}\ \bibnamefont {Sun}},\ }\href
  {\doibase 10.1103/PhysRevB.91.214416} {\bibfield  {journal} {\bibinfo
  {journal} {Phys. Rev. B}\ }\textbf {\bibinfo {volume} {91}},\ \bibinfo
  {pages} {214416} (\bibinfo {year} {2015})}\BibitemShut {NoStop}%
\bibitem [{\citenamefont {Pai}\ \emph {et~al.}(2015)\citenamefont {Pai},
  \citenamefont {Ou}, \citenamefont {Vilela-Le\~ao}, \citenamefont {Ralph},\
  and\ \citenamefont {Buhrman}}]{PaiPRB2015}%
  \BibitemOpen
  \bibfield  {author} {\bibinfo {author} {\bibfnamefont {C.-F.}\ \bibnamefont
  {Pai}}, \bibinfo {author} {\bibfnamefont {Y.}~\bibnamefont {Ou}}, \bibinfo
  {author} {\bibfnamefont {L.~H.}\ \bibnamefont {Vilela-Le\~ao}}, \bibinfo
  {author} {\bibfnamefont {D.~C.}\ \bibnamefont {Ralph}}, \ and\ \bibinfo
  {author} {\bibfnamefont {R.~A.}\ \bibnamefont {Buhrman}},\ }\href {\doibase
  10.1103/PhysRevB.92.064426} {\bibfield  {journal} {\bibinfo  {journal} {Phys.
  Rev. B}\ }\textbf {\bibinfo {volume} {92}},\ \bibinfo {pages} {064426}
  (\bibinfo {year} {2015})}\BibitemShut {NoStop}%
\bibitem [{\citenamefont {Berger}\ \emph {et~al.}(2018)\citenamefont {Berger},
  \citenamefont {Edwards}, \citenamefont {Nembach}, \citenamefont {Karenowska},
  \citenamefont {Weiler},\ and\ \citenamefont {Silva}}]{BergerPRB2018}%
  \BibitemOpen
  \bibfield  {author} {\bibinfo {author} {\bibfnamefont {A.~J.}\ \bibnamefont
  {Berger}}, \bibinfo {author} {\bibfnamefont {E.~R.~J.}\ \bibnamefont
  {Edwards}}, \bibinfo {author} {\bibfnamefont {H.~T.}\ \bibnamefont
  {Nembach}}, \bibinfo {author} {\bibfnamefont {A.~D.}\ \bibnamefont
  {Karenowska}}, \bibinfo {author} {\bibfnamefont {M.}~\bibnamefont {Weiler}},
  \ and\ \bibinfo {author} {\bibfnamefont {T.~J.}\ \bibnamefont {Silva}},\
  }\href {\doibase 10.1103/PhysRevB.97.094407} {\bibfield  {journal} {\bibinfo
  {journal} {Phys. Rev. B}\ }\textbf {\bibinfo {volume} {97}},\ \bibinfo
  {pages} {094407} (\bibinfo {year} {2018})}\BibitemShut {NoStop}%
\end{thebibliography}
\end{document}